\documentclass[
nofootinbib,showpacs,preprintnumbers,amsmath,amssymb]{revtex4}

\usepackage{graphicx}
\usepackage{dcolumn}
\usepackage{oldgerm}
\usepackage{amsmath}
\usepackage{psfrag}
\usepackage{color}
\usepackage{latexsym,amsfonts}
\usepackage{graphicx}
\newcommand{\bee}{\begin{eqnarray}}
\newcommand{\eee}{\end{eqnarray}}
\newcommand{\be}{\begin{equation}}
\newcommand{\ee}{\end{equation}}
\newcommand{\ds}{\displaystyle}

\begin{document}

\title{The elastic electron-deuteron scattering beyond one-photon exchange}
\author{A.P.~Kobushkin}
 \email{kobushkin@bitp.kiev.ua}
\affiliation{%
Bogolyubov Institute for Theoretical Physics, Metrologicheskaya Street, 14B, 03680 Kiev, Ukraine
}%
\author{Ya.D.~Krivenko-Emetov} 
\affiliation{%
Institute for Nuclear Research, Prospekt Nauki 47, 03680 Kiev, Ukraine
}%
\author{S.~Dubni\v cka}
\affiliation{%
Institute of Physics, Slovak Academy of Sciences, Bratislava, Slovak Republic
}%

\date{\today}

\begin{abstract}
We discuss elastic $ed$ scattering beyond the Born approximation. The reaction amplitude contains six generalized form factors, but only three linearly independent combinations of them (
generalized charge, quadrupole and magnetic form factors) contribute to the reaction cross section in second order perturbation theory. We examine the two-photon exchange and find that it includes two types of diagrams, where two virtual photons are interacting with the same nucleon and where the photons are interacting with different nucleons. It is shown that the two-photon exchange amplitude is strongly connected with the deuteron wave function at short distances.
\end{abstract}

\maketitle

\section{\label{sec:Introduction}Introduction}
The study of electron scattering on the nucleon and the light nuclei provides a convenient tool to study the 
structure of strongly interacting systems. Due to the smallness of the fine structure constant 
$\alpha \approx \frac1{137}$, one may expect that the Born approximation (one-photon exchange, OPE) 
should describe such processes with an accuracy of a few percent. Nevertheless, JLab polarization 
measurements of $G_E^p(Q^2)/G_M^p(Q^2)$ \cite{Jlab.Jones,Jlab.Gayou,Punjabi}, together with their 
theoretical analysis \cite{Guichon,Blunden,BorisyukKob,BorisyukKob_phen}, show that higher order 
perturbative effects, such as two-photon exchange (TPE), can strongly affect some observables 
of the elastic electron-nucleon scattering.

Also for more complicated hadronic systems, like the deuteron, $^3$He,  $^4$He, etc., TPE should contribute.
Thus for precise studies of these nuclei a quantitative theoretical investigation of TPE effects is important; 
until now only a few estimates have been done of the contribution of TPE \cite{DeForest,Gunion,Franco,
Boitsov,Lev,DongChen}.

The aim of this paper is to estimate the TPE amplitude for $ed$-scattering in the framework of 
semi-relativistic calculations, with deuteron wave functions from ``realistic'' NN~potentials.

The paper is organized as follows. In Sect.~\ref{sec:Kinem} we study the general structure of the 
reaction amplitude beyond OPE and define six independent generalized form factors which determine 
the amplitude. We show that only three linearly independent combinations of these generalized form factors
contribute to the cross section in second order perturbation theory. We call the corresponding 
combinations of the form factors generalized charge, quadrupole and magnetic form factors. These 
generalized form factors 
are computed in Sect.~\ref{sec:two-photon}. Sect.~\ref{sec:num} contains numerical results and a 
brief discussion.
\section{Kinematics and definitions \label{sec:Kinem}}
The electron and deuteron momenta in the initial and final states of elastic $ed$ scattering are denoted 
by $k$, $k'$ and $d$, $d'$, respectively; $q=k-k'$ is the transferred momentum; $M$ and $m$ are deuteron 
and nucleon masses; actual calculations will be done with $m\approx\frac12 M$.

\begin{figure}[t]
  \centering  
  \includegraphics[height=0.2\textheight]{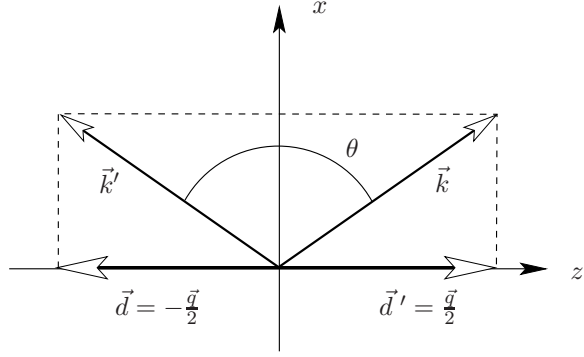}
  \caption{The electron and deuteron 3-momenta in the Breit frame.\label{fig:BreitF}}
\end{figure}

All calculations are done in the Breit frame, where the deuteron has the same energy $E_d$ in the initial and final state and moves along the $z$-direction (Fig.~\ref{fig:BreitF}). We get
\begin{gather} 
d_0=d'_{0}=E_d=\sqrt{M^2+Q^2/4}, \nonumber\\
\vec d_\perp=\vec d\,'_\perp=0, \qquad d_3=-d'_{3}=-Q/2,\nonumber\\
 q_0=q_1=q_2=0,\qquad q_3=Q,\\
k_0=k'_0\equiv E_e, \qquad \vec k_\perp=\vec k\,'_\perp, \qquad k_3=-k'_3=Q/2, \nonumber
\label{Breitframe}
\end{gather}
where $Q$ is the modulus of the transferred momentum.
For definiteness we will assume that the transverse momentum of the electron is directed along the $x$-axis
\begin{equation}
\label{Breitframe0}
k_1=E_e\cos\tfrac{\theta}2, \qquad k_2=0, \qquad k_3=\tfrac12 Q=E_e\sin\tfrac{\theta}2.
\end{equation}
In this frame the commonly used polarization parameter $\epsilon$ can be expressed in terms of the electron scattering angle $\theta$ by
\be
\epsilon=\frac{\cos^2\tfrac{\theta}2}{1+\sin^2\tfrac{\theta}2};
\ee
note that $\mathrm{tg}^2\textstyle{\frac{\theta}2}=(1+\eta)\mathrm{tg}^2\textstyle{\frac{\theta_\mathrm{lab}}2}$, where $\eta=\frac {Q^2}{4M^2}$.

The polarization vectors for the incoming and outgoing deuterons with spin $z$-projection $\lambda$ are denoted by $\epsilon_{(\lambda)}(d)$ and $\epsilon_{(\lambda)}(d')$, respectively,
\begin{equation}
\label{PolarizationVec}
\begin{split}
\epsilon_{(\pm 1)}(d)&=\epsilon_{(\pm 1)}(d')=-\sqrt{\tfrac12}(0,\pm1,i,0),\\
\epsilon_{(0)}(d)&=\frac1M\left(-\frac{Q}2,0,0,E_d\right), \\ 
\epsilon_{(0)}(d')&=\frac1M\left(\frac{Q}2,0,0,E_d\right).
\end{split}
\end{equation} 
The required electron electromagnetic current $j^\mu=\bar u_h(k')\gamma^\mu u_h(k)$ is written as: 
\begin{equation}
\label{current1}
\begin{aligned}
&j_0=2E_e\cos\tfrac{\theta}2, \qquad &&j_1=-2E_e,\\ 
&j_2=-2ihE_e\sin\tfrac{\theta}2, &&j_3=0.
\end{aligned}
\end{equation}
Because the electrons are ultra-relativistic, the helicities of the incoming and outgoing electrons 
are the same; its sign will be specified by $h$.

Instead of the usual reaction amplitude $\mathcal M$ it is useful to introduce the reduced amplitude $T_{\lambda'\lambda,h}$ by
\be \label{Gen.Amplitue}
\mathcal M=\frac{16\pi\alpha}{Q^2}E_eE_d T_{\lambda'\lambda,h}.
\ee

It follows from $P$ and $T$ invariance that, in the Breit frame, this amplitude must have the following properties
\be\label{space-time}
\begin{array}{ll}
T_{\lambda'\lambda; h}=(-1)^{\lambda-\lambda'}T_{-\lambda'-\lambda; -h},&\text{from $P$ invariance},\\
T_{\lambda'\lambda; h}=T_{-\lambda-\lambda'; h},&\text{from $T$ invariance}.
\end{array}
\ee
The reaction amplitude is determined by six independent invariant amplitudes (form factors), which are specified
by the following parametrization:
\be\label{Gen.Reduce_Amplitue}
T_{\lambda'\lambda; h}=\left(
\begin{array}{ccc}
\mathcal G_{11}\cos\frac{\theta}2&-\sqrt\frac{\eta}{2}\mathcal G_{10}^{h} & \mathcal G_{1,-1}^{h}\\
\sqrt\frac{\eta}{2}\mathcal G_{10}^{-h} &\mathcal G_{00}\cos\frac{\theta}2&-\sqrt\frac{\eta}{2}\mathcal G_{10}^{h}\\
\mathcal G_{1,-1}^{-h} & \sqrt\frac{\eta}{2}\mathcal G_{10}^{-h}& \mathcal G_{11}\cos\frac{\theta}2
\end{array}
\right),
\ee
where lines and columns correspond to the allowing order: $(\lambda',\lambda)=+1,0,-1$, and
\be \label{Gen.Gmn}
\mathcal G_{10}^{h}=f_1+h\sin\tfrac{\theta}2 f_2, \qquad
\mathcal G_{1,-1}^{h}=f_3+h\sin\tfrac{\theta}2 f_4.
\ee
The form factors $\mathcal G_{11}$, $\mathcal G_{00}$, $f_1$, ..., $f_4$ are complex functions of the two independent kinematical variables, for example $Q^2$ and $\theta$.

The relation of the amplitude $T_{\lambda'\lambda; h}$ to the invariant amplitudes $G_1,...,G_6$ used by other authors \cite{GakhTomasi,DongPRC} is given in Appendix~\ref{AppendixOPE}.

Later on the amplitude (\ref{Gen.Reduce_Amplitue}) will be expanded in $\alpha$, and only terms of order zero 
and one will be kept. As shown in Eq.~(\ref{Gi_ampl_0}), at zero order (OPE approximation) the amplitude 
written in terms of the charge, magnetic and quadruple form factors ($G_C$, $G_M$ and $G_Q$) becomes
\begin{widetext}
\be\label{Gen.Reduce_Amplitue.ONP}
T_{\lambda'\lambda; h}^{(0)}=\left(
\begin{array}{ccc}
\left(G_C-\frac23\eta G_Q\right)\cos\tfrac{\theta}2 & -\sqrt\frac{\eta}{2}G_M(1+h\sin\tfrac{\theta}2) & 0\\
\sqrt\frac{\eta}{2}G_M(1-h\sin\tfrac{\theta}2) &\left( G_C+\frac43\eta G_Q\right)\cos\tfrac{\theta}2 &
-\sqrt\frac{\eta}{2}G_M(1+h\sin\tfrac{\theta}2)\\
0 & \sqrt\frac{\eta}{2}G_M(1-h\sin\tfrac{\theta}2)& \left(G_C-\frac23\eta G_Q\right)\cos\tfrac{\theta}2
\end{array}
\right),
\ee
\end{widetext}
where the form factors $G_C(Q^2)$, $G_Q(Q^2)$ and $G_M(Q^2)$ are real and depend upon $Q^2$ only.

Next the generalized electric, quadrupole and magnetic  from factors $\mathcal G_C(Q^2,\theta)$, 
$\mathcal G_Q(Q^2,\theta)$, $\mathcal G_M(Q^2,\theta)$, which reproduce the spin structure of 
Eq.~(\ref{Gen.Reduce_Amplitue.ONP}), and the additional form factors $g_1(Q^2,\theta)$, 
$g_2(Q^2,\theta)$, $g_3(Q^2,\theta)$ are introduced as follows:
\be \label{GcGqGm}
\begin{split}
&\mathcal G_{11}=\mathcal G_C-\tfrac23\eta \mathcal G_Q,\qquad
\mathcal G_{00}=\mathcal G_C+\tfrac43\eta \mathcal G_Q,\\
&f_1=\mathcal G_M + g_1\sin^2\tfrac{\theta}2,\qquad f_2=\mathcal G_M - g_1,\\
&f_3=g_2,\qquad f_4=g_3.
\end{split}
\ee

In OPE+TPE approximation the form factors can be written as:
\be \label{FFs_expan}
\begin{split}
&\mathcal G_C=G_C+\delta\mathcal G_C,\qquad \mathcal G_Q=G_Q+\delta\mathcal G_Q, \\ &\mathcal G_M=G_M+\delta \mathcal G_M,
\end{split}
\ee
where $\delta$ stands for the terms of order $\alpha$; likewise, the form factors $g_{1,2,3}$ are also 
proportional to $\alpha$.

By standard calculation one derives the differential cross section
\be\label{Cross-s}
\dfrac{d\sigma}{d\Omega}=\dfrac{\sigma_{\mathrm M}}{\cos^2\tfrac{\theta}2}\overline{|T|^2},
\ee
where $\sigma_{\mathrm M}$ is the Mott cross section and
\be \label{Gen.cross-section}
\begin{split}
&\overline{|T|^2}=\tfrac16\sum_{\lambda,\lambda',h}|T_{\lambda'\lambda; h}|^2=\\
&=\cos^2\tfrac{\theta}2 \left[\mathcal A(Q^2,\theta) + \mathrm{tg}^2\tfrac{\theta_\mathrm{lab}}2 \mathcal B(Q^2,\theta)\right]+\mathcal O(\alpha^2)=\\
&=\left(1+\sin^2\tfrac{\theta}2\right)\left[\epsilon |\mathcal G_E(Q^2,\theta)|^2 + \frac{2}{3}\eta |\mathcal G_M(Q^2,\theta)|^2 \right]+\\ 
&\hspace{2.cm}+\mathcal O(\alpha^2),
\end{split}
\ee
with
\be \label{Gen.A_and_B}
\begin{split}
&\mathcal A(Q^2,\theta)=|\mathcal G_C(Q^2,\theta)|^2+\tfrac89\eta^2|\mathcal G_Q(Q^2,\theta)|^2 +\\
&\hspace{1.5cm}+\tfrac23\eta|\mathcal G_M(Q^2,\theta)|^2, 
\\
&\mathcal B(Q^2,\theta)=\tfrac43 (1+\eta)\eta|\mathcal G_M(Q^2,\theta)|^2,\\
&\mathcal G_E^2=|\mathcal G_C(Q^2,\theta)|^2+\tfrac89\eta^2|\mathcal G_Q(Q^2,\theta)|^2.
\end{split}
\ee
The advantage of using the form factors $\mathcal G_C$, $\mathcal G_Q$ and $\mathcal G_M$ is that the 
expression for the cross section has the same form as the Rosenbluth formula; nevertheless the Rosenbluth 
separation of the structure functions $\mathcal A(Q^2,\theta)$ and $\mathcal B(Q^2,\theta)$ can no longer 
be done because they depend on two variables.
\section{\label{sec:two-photon}Calculation of the two-photon exchange}
\begin{figure}
  \centering 
\includegraphics[height=0.325\textheight]{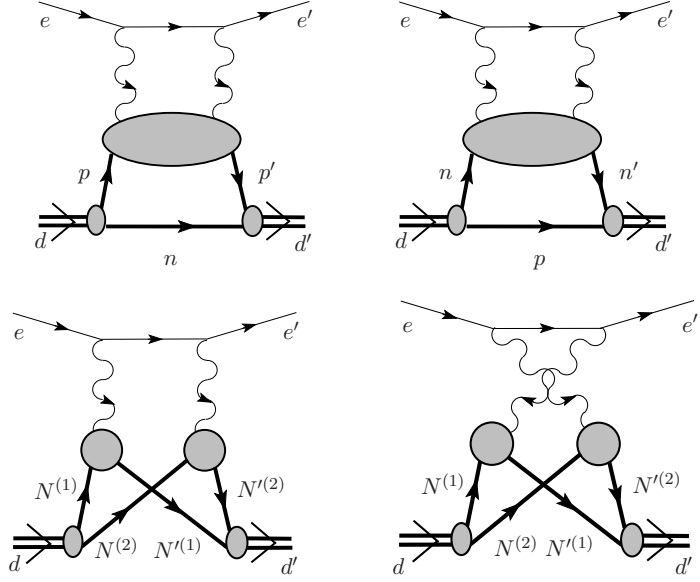}
  \caption{Two-photon exchange diagrams. The top diagrams correspond to the amplitudes $\mathcal M_p^{\mathrm I}$ and $\mathcal M_n^{\mathrm I}$, the bottom diagrams to the amplitudes $\mathcal M_{\mathrm P}^{\mathrm{II}}$ (left) and $\mathcal M_{\mathrm X}^{\mathrm{II}}$ (right).}
 \label{fig:2photon}
\end{figure}

In what follows the contribution of meson exchange currents to the TPE amplitude will be neglected, and 
two types of TPE diagrams will be considered, where the virtual photons interact directly with the nucleons
\be \label{TPE-definition}
\mathcal M_2=\mathcal M^{\mathrm{I}}+\mathcal M^{\mathrm{II}}.
\ee
One of them, $\mathcal M^{\mathrm{I}}=\mathcal M^{\mathrm{I}}_p+\mathcal M^{\mathrm{I}}_n$, corresponds to diagrams, where both photons interact with the same nucleon (Fig.~\ref{fig:2photon}, top). The other type, $\mathcal M^{\mathrm{II}}=\mathcal M^{\mathrm{II}}_{\mathrm P}+\mathcal M^{\mathrm{II}}_{\mathrm X}$, corresponds to the diagrams, where the photons interact with different nucleons (Fig.~\ref{fig:2photon}, bottom).

The deuteron structure will be described by the non-relativistic wave function
\bee
&\Psi(\lambda,\vec{p}\,)=\sum_{\sigma_1,\sigma_2}\Psi_{\sigma_1\sigma_2}(\lambda,\vec{p}\,)=\nonumber\\
&=\sum_{\sigma_1,\sigma_2}\left[\sqrt{\textstyle\frac{1}{4\pi}}\left\langle{\textstyle\frac12\frac12}\sigma_1\sigma_2|1\lambda\right\rangle
 U_0(p)-\right.\label{WaveFunction} \\
&\left.
-\sum_{\xi,M}Y_{2\xi}(\widehat{p})\left\langle{\textstyle\frac12\frac12}\sigma_1 \sigma_2|1M\right\rangle\langle21\xi M|1\lambda\rangle U_2(p)\right]\times\nonumber\\
&\times |N_1\sigma_1,N_2\sigma_2\rangle,\nonumber
\eee
where $\vec p$ is the internal momentum in the deuteron, $|N_1\sigma_1,N_2\sigma_2\rangle$ is the 
spin-isospin wave function of the two nucleons and $\left\langle...|...\right\rangle  $ 
are Clebsh-Gordan coefficients.
\subsection{$\mathcal M^{\mathrm{I}}_N$ diagram}
The TPE amplitude for a nucleon $N$ has the following structure \cite{Guichon}
\be\label{TPE_N}
\mathcal{M}_{2\gamma N}=\frac{4\pi\alpha}{Q^2}\bar u'_h\gamma_\mu u_h \left\langle \vec p{\,'}_{\!\!\!N}\sigma'\left|\widehat H^\mu_N \right|\vec p_N\,\sigma\right\rangle ,
\ee
where $\widehat H^\mu_N$ is the ``effective hadron current''
\be
\widehat H^\mu_N=\Delta\widetilde F_{1}^N\gamma^\mu -\Delta\widetilde F_{2}^N[\gamma^\mu,\gamma^\nu]\frac{q_\nu}{4m}+
\widetilde F_{3}^NK_\nu\gamma^\nu\frac{P^\mu}{m^2}.
\label{TPE.I.3}
\ee
In Eqs.~(\ref{TPE_N}) and (\ref{TPE.I.3}) $p_N$ and $p{\,'}_{\!\!\!N}$ are the nucleon momenta, $\sigma$ and $\sigma'$ are the nucleon spin projections, $\left|\vec p_N\,\sigma\right\rangle $ and $\left|\vec p{\,'}_{\!\!\!N}\,\sigma'\right\rangle $ are the nucleon spinors, $K=(k+k')/2$, $P=(p_N+p{\,'}_{\!\!\!N})/2$; $\Delta\widetilde F_{1}^N$ and $\Delta\widetilde F_{2}^N$ may be called corrections to the Dirac and Pauli form factors and $\widetilde F_{3}^N$ is a new form factor. All the 
quantities $\Delta\widetilde F_{1}^N$, $\Delta\widetilde F_{2}^N$ and $\widetilde F_{3}^N$ are of order $\alpha$. They are complex functions of two kinematical variables, e.g. $Q^2$ and $\nu=4PK$.

Because we employ a non-relativistic deuteron wave function we have to put
\be\label{N_momenta}\begin{split}
&p_p\approx (m, \tfrac12 \vec d +\vec p\,),\qquad p_n\approx (m, \tfrac12 \vec d -\vec p\,),\\
&p'_p\approx (m, \tfrac12 \vec d{\,'} +\vec p{\,'}),\qquad p'_n\approx (m, \tfrac12 \vec d{\,'} -\vec p{\,'}),
\end{split}
\ee
where $\vec p$ and $\vec p{\,'}$ are the internal momenta in the deuteron.

From (\ref{TPE_N}) it follows that the $\mathcal M^{\mathrm{I}}$ amplitude is given by
\be\label{TPE.I.3.a}
\mathcal M^{\mathrm{I}}=\frac{4\pi\alpha}{Q^2}\bar u'\gamma_\mu u \mathcal D^\mu(\lambda',\lambda),
\ee
where $\mathcal D^\mu(\lambda'\lambda)$ is an effective deuteron current. The latter is derived in the same way as the deuteron current $J^\mu$ in the impulse approximation \cite{Gourdin} with the nucleon current substituted by the effective hadron current (\ref{TPE.I.3}),
\be\label{TPE.I.3.b}
\mathcal D^\mu(\lambda',\lambda)=\frac{E_d}{m}\int d^3p \Psi^\dag(\vec p+\tfrac12\vec q,\lambda')(\widehat H^\mu_p+\widehat H^\mu_n)\Psi(\vec p,\lambda).
\ee
Here $\Psi(\vec p,\lambda)$ and $\Psi(\vec p+\tfrac12 \vec q,\lambda')$ are wave functions of the deuteron in the initial and final states\footnote{Here the additional multiplier $(2m)^{-1}$ appears due to the additional multiplier $2m$ in the effective current [see Eq.~(\ref{hadron_vector})] in comparison with the nucleon current (3) and (4) of Ref.~\cite{Gourdin}.}.

Later on we will need a non-relativistic reduction of matrix elements of the effective hadron current. Retaining the terms linear in the nucleon momentum one gets (see Appendix~\ref{App.TPE.I.3})
\be\label{hadron_vector}\begin{split}
&\left\langle \vec p{\,'}_{\!\!\!N}\sigma'\left|\widehat H^0_N \right|\vec p_N\,\sigma\right\rangle  \approx\\ 
&\approx 2m\chi^\dag_{\sigma'}\left(
\delta \mathcal G_E^N-\frac{iE_eQ\sigma^2}{2m^2}\cos\tfrac{\theta}2\widetilde F_3^N
\right)\chi_\sigma \equiv 
\chi^\dag_{\sigma'}\mathcal H^0_N\chi_{\sigma},\\
&\left\langle \vec p{\,'}_{\!\!\!N}\sigma'\left| \widehat {\vec H}_N \right|\vec p_N\,\sigma\right\rangle \approx\\
&\approx  \chi^\dag_{\sigma'}\left[
i \left(\vec \sigma\times\vec q\,\right) \left(\delta \mathcal G_M^N -\frac{\epsilon E_e}{m}\widetilde F_3^N\right)+ 2\vec P \delta \mathcal G_E^N 
\right]\chi_m\equiv \\
&\equiv\chi^\dag_{\sigma'}\vec{\mathcal H}_N\chi_{\sigma},
\end{split}
\ee
where $\vec \sigma$ are Pauli matrices and $\chi_{\sigma'}$, $\chi_{\sigma}$ are Pauli spinors.

The generalized nucleon electric and magnetic form factors are defined by (see Ref.~\cite{BorisyukKob_phen})
\be
\begin{split}\label{Corrections}
&\delta \mathcal G_E^N= \Delta\widetilde F_1^N -\tau \Delta\widetilde F_2^N+\frac{\nu}{4m^2}\widetilde F_3^N,\\
&\delta \mathcal G_M^N=\Delta\widetilde F_1^N + \Delta\widetilde F_2^N+\frac{\epsilon\nu}{4m^2}\widetilde F_3^N.
\end{split}
\ee
Here $\tau = \frac{Q^2}{4m^2}\approx 4\eta$ and $\nu\approx m E_e$. After substitution of 
Eqs.~(\ref{hadron_vector}) and (\ref{Corrections}) into Eq.~(\ref{TPE.I.3.b}) the matrix elements of the effective deuteron current between the initial and final deuteron states become
\be
\begin{split}
&\mathcal D^0(\lambda',\lambda)=\left\{
\begin{array}{lll}
2E_d\left(\delta \mathcal G_C^{\mathrm I}-\frac23\eta \delta \mathcal G_Q^{\mathrm I}\right),&\text{if}&\lambda =\lambda'=\pm 1,\\
2E_d\left(\delta \mathcal G_C^{\mathrm I}+\frac43\eta \delta \mathcal G_Q^{\mathrm I}\right),&\text{if}&\lambda =\lambda'=0,\\
-2i\frac{E_eE_d}{m}\times\\
\times\sqrt{\eta} \cos\frac{\theta}2\langle\lambda' |J_2|\lambda\rangle \mathcal F_3,&\text{if},&\lambda'-\lambda =\pm1,\\
0,&\text{if}&\lambda'-\lambda =\pm2,
\end{array}
\right.\\
&\vec{\mathcal D}(\lambda',\lambda)=i\langle \lambda'|\vec J\times \vec q\,|\lambda\rangle\frac{E_d}{M}
\left(\delta \mathcal G_M^{\mathrm I}-\frac{\epsilon E_e}{m} \mathcal F_3\right),
\end{split}
\ee
where $\vec J=(J_1,J_2,J_3)$ is an operator of the deuteron total angular momentum, and
\be\label{TPE.I.5}
\begin{split}
&\delta \mathcal G_C^{\mathrm I} = 2\delta\mathcal G_E^S\left[I_{00}^0(Q)+I_{22}^0(Q)\right],\\
&\delta \mathcal G_Q^{\mathrm I} = \frac{3\sqrt 2}{\eta}\delta\mathcal G_E^S\left[I_{20}^2(Q)-\frac1{2\sqrt 2}I_{22}^2(Q)\right],\\
&\delta \mathcal G_M^{\mathrm I} = \frac{M}m\left\{\frac32\delta\mathcal G_E^S\left[I_{22}^0(Q)+I_{22}^2(Q)\right]+\right.\\
 & +\left.
2\delta\mathcal G_M^S\left[I_{00}^0(Q)-\tfrac12 I_{22}^0(Q)+\sqrt{\tfrac12} I_{20}^2(Q)+\tfrac12 I_{22}^2(Q)\right]\right\},\\
&\mathcal F_3 =2\frac{M}{m}\widetilde F_3^S\left[I_{00}^0(Q)-\tfrac12 I_{22}^0(Q)+\sqrt{\tfrac12} I_{20}^2(Q)+\right.\\
&\hspace{2.cm}\left.+\tfrac12 I_{22}^2(Q)\right].
\end{split}
\ee
In these expressions the notation  
$I_{\ell'\ell}^{L}(Q)=\int_0^\infty dr j_L\left(\tfrac{1}{2}Qr\right)u_{\ell'}(r)u_{\ell}(r)$ was used, 
where $j_L(x)$ is a spherical Bessel function, $u_\ell(r)$ is the radial deuteron wave function for orbital momentum $\ell$ and $\delta\mathcal G_E^S=\frac12(\delta\mathcal G_E^p + \delta\mathcal G_E^n)$, etc. Contracting the effective deuteron current with the electron current $j_\mu$ one arrives at
\be \label{G4I}
g_1^\mathrm{I}=-\epsilon\frac{E_e}{m}\mathcal F_3.
\ee
The corrections $g_2$ and $g_3$ are obviously vanishing for the first type of diagrams, $g_2^\mathrm{I}=g_3^\mathrm{I}=0$.
\subsection{$\mathcal M^{\mathrm{II}}$ diagrams }
In Ref.~\cite{Lev} the contribution of the $\mathcal M^{\mathrm{II}}$ diagram was estimated within a 
non-relativistic approach with a Gaussian deuteron wave function. The present calculations are similar, but 
use a deuteron wave functions extracted from a  ``realistic'' NN~potentials; also, a modern parametrization for the nucleon form factors has been adopted.

The appropriate amplitude is given by the sum of two diagrams displayed at the bottom of Fig.~\ref{fig:2photon}, $\mathcal{M}^{\mathrm{II}}=\mathcal{M}^{\mathrm{II}}_{\mathrm{P}}+\mathcal{M}^{\mathrm{II}}_{\mathrm{X}}$, where
\begin{equation}
i\mathcal{M}^{\mathrm{II}}_{\mathrm{P,X}}=\int \frac{d^4p}{(2\pi)^4}\frac{d^4p\,'}{(2\pi)^4}\
\widetilde t^{\;\mathrm{P,X}}_{\mu\nu}G(\Delta_1,\Delta_2)\frac{{\mathcal{T}}_\mathrm{P,X}^{\mu\nu(\lambda'\lambda)}}{D}\;.
\label{TPE.2}
\end{equation}
Here $p=\frac12(p^{(1)}-p^{(2)})$ and  $p'=\frac12(p'^{(1)}-p'^{(2)})$ are the relative momenta in the initial and final deuteron,
\be
\begin{split}
&\widetilde t^{\;\mathrm P}_{\mu\nu}=\displaystyle{\frac{\bar u_{h}(k')\left(-ie\gamma_\mu\right)i(l\!\!\!/ + \mu)\left(-ie\gamma_\nu\right)u_h(k)}{l^2-\mu^2+i0}}\;,\\
&\widetilde  t^{\;\mathrm X}_{\mu\nu}=\widetilde t^{\;\mathrm P}_{\nu\mu}\;  ,\\
&G(\Delta_1,\Delta_2)=\displaystyle{\frac{ -i}{\Delta_1^2-\kappa^2+i0}\cdot\frac{ -i}{\Delta_2^2-\kappa^2+i0}},\\
& \mathcal{T}^{\mu\nu(\lambda'\lambda)}_\mathrm{P}\left(\Delta_1^2,\Delta_2^2\right)=\\
&\hspace{0.35cm}=(ie)^2\mathrm{Tr}\left\{
id^{(\lambda')}(p'^{(1)},p'^{(2)})\;
(p\!\!\!/'^{(1)}+m)\times\right.\\
&\hspace{0.35cm}\left.\times\Gamma^\mu_1\left(\Delta_1^2\right)(p\!\!\!/^{(1)}+m)id^{(\lambda)}(p^{(1)},p^{(2)})
(p\!\!\!/^{(2)}-m)\times
\right.\\
& \hspace{0.35cm}\left.\times \bar\Gamma^\nu_2\left(\Delta_2^2\right)\;(p\!\!\!/'^{(2)}-m)
\right\},\\
& \mathcal{T}^{\mu\nu(\lambda'\lambda)}_\mathrm{X}\left(\Delta_1^2,\Delta_2^2\right)=\mathcal{T}^{\mu\nu(\lambda'\lambda)}_\mathrm{P}\left(\Delta_2^2,\Delta_1^2\right),\\
&D=[(p'^{(1)})^2-m^2+i0][(p^{(1)})^2-m^2+i0]\times\\
&\hspace{0.35cm}\times[(p^{(2)})^2-m^2+i0][(p'^{(2)})^2-m^2+i0].
\label{Nucleon_propagator}
\end{split}
\ee
where P and X superscripts (subscripts) mean appropriate quantities related to diagrams with ``parallel'' photons (left bottom, Fig.2) and ``crossed' photons (right bottom, Fig.2); $A\!\!\!/ \equiv
  A_\mu\gamma^\mu$.
In (\ref{Nucleon_propagator}) we use the following notations, $l$ is the 4-momentum of the intermediate electron, $\Delta_1=k-l$ and $\Delta_2=l-k'$ are the 4-momenta of the virtual photons and $\kappa$ is an infinitesimal photon mass introduced in the photon propagators to regulate the infrared divergences; $\Gamma^\mu_1(\Delta_1^2)$ and $\bar\Gamma^\mu_2(\Delta_2^2)$ are electromagnetic currents for the nucleon and anti-nucleon, in which the form factors are functions of $\Delta_1^2$ and $\Delta_2^2$, respectively; $d^{(\lambda)}(p^{(1)},p^{(2)})$ and $d^{(\lambda')}(p'^{(1)},p'^{(2)})$ are $dpn$ vertex functions for the initial and final deuteron.

In expression (\ref{Nucleon_propagator}) for 
$\mathcal{T}^{\mu\nu(\lambda'\lambda)}_\mathrm{P,X}$ moving along the 
nucleon loop (bold lines in the bottom diagrams in Fig.~\ref{fig:2photon}): 
a line with an arrow opposite to the motion corresponds to a fermion 
propagator and a line with an arrow along the motion corresponds to an 
anti-fermion propagator.
\begin{figure*}[tb]
  \centering 
  \includegraphics[height=0.11\textheight]{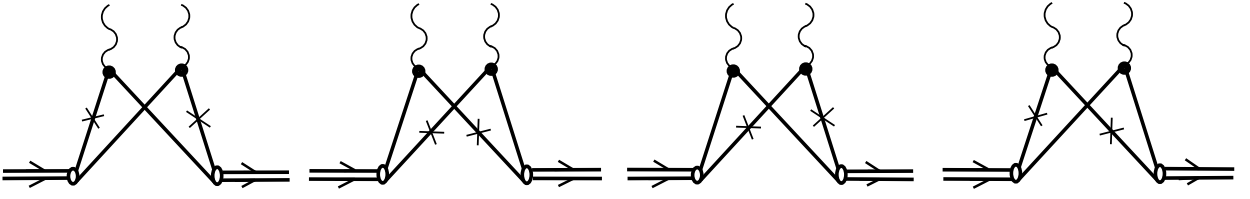}
  \caption{Four types of poles taken into account in integration over $dp_0$ and $dp_0'$.\label{fig:Poles}}
 \end{figure*}

An infrared divergent term appears in $\mathcal{M}^{\mathrm{II}}$ when one photon is soft, $\Delta_1\to 0$, $\Delta_2\to q$ or $\Delta_1\to q$, $\Delta_2\to 0$. It is canceled by radiative corrections which are not of interest 
in this paper. The configuration where each intermediate photon carries 
about half of the transferred momentum (hard-photon approximation) is 
emphasized,
\begin{equation}
\Delta_1\sim \Delta_2\sim \frac{q}2\ .
\label{2photonConfig}
\end{equation}
In this case there is no infrared divergent terms.

To relate the $dpn$ vertex to the deuteron wave function with one of the nucleons 
on the mass shell one has to integrate over  $dp_0$ and $dp_0'$. Four types 
of poles contribute to this integral (see Fig.~\ref{fig:Poles}). What follows is a 
 discussion of the contribution coming from the poles of the first diagram 
of Fig.~\ref{fig:Poles}
\begin{equation}\label{Pole_int}
\begin{split}
&dT^{\mu\nu}_\mathrm{P,X} \equiv \frac{d^4pd^4p'}{(2\pi)^8}\frac{\mathcal{T}_\mathrm{P,X}^{\mu\nu(\lambda'\lambda)}}{ D} \to-\frac14\frac{d^3pd^3p'}{2E_12E'_2(2\pi)^6}\times\\
&\times\frac{\mathcal{T}_\mathrm{P,X}^{\mu\nu(\lambda'\lambda)}}{\left({p'^{(1)}}^2-m^2+i0 \right)\left({p^{(2)}}^2-m^2+i0\right)},
\end{split}
\end{equation}
where $E_1=\sqrt{m^2+\left(\vec p-\tfrac14\vec q\;\right)^2}$ and $E_2'=\sqrt{m^2+\left(\vec p\,'-\tfrac14\vec q\;\right)^2}$.

One can use the expansion
\begin{equation}\label{Expansion1}
\begin{split}
&p\!\!\!/\,^{(1)}+m=\sum_{\sigma_1}|\vec p\,^{(1)}\sigma_1\rangle\langle\vec p\,^{\,(1)}\sigma_1|,
\\
&p\!\!\!/\,'^{(2)}-m=\sum_{\sigma'_2} |\vec p\,'^{(2)}\sigma'_2;c\rangle \langle \vec p\,'^{(2)}\sigma'_2;c|
\end{split}
\end{equation}
(in the last equation $c$ means charge conjugated spinor) and define the wave functions of the initial and final deuteron by
\begin{equation}
\begin{split}
\label{RelWaveFunction} 
&\phi^{(\lambda)}(p^{(1)},p^{(2)})=
\frac{d^{(\lambda)}\left(p^{(1)},p^{(2)}\right)}{{p^{(2)}}^2-m^2+i0}\,,
\\
&\phi^{(\lambda')}(p'^{(1)},p'^{(2)})=
\frac{d^{(\lambda')}\left(p'^{(1)},p'^{(2)}\right)}{{{p'^{(1)}}^2-m^2+i0}}\,. 
\end{split}
\end{equation}
These wave functions are normalized by the condition
\be\label{WFNormalization}
\int\dfrac{d^3p}{2E_1(2\pi)^3}\mathrm{Tr}\;\phi^{\dag\,(\lambda)}(p^{(1)},p^{(2)})\phi^{(\lambda)}(p^{(1)},p^{(2)})=1
\ee
[and similarly for $\phi^{(\lambda')}(p'^{(1)},p'^{(2)})$], which comes from the requirement $G_C(0)=1$.

Note that in general the nucleons $N'^{(1)}$ and $N^{(2)}$ are not on-shell and at this step one cannot use expansions similar to (\ref{Expansion1}) for $p\!\!\!/'\,^{(1)}+m$ and $p\!\!\!/\,^{(2)}-m$. Nevertheless we will assume that the relative momenta in the initial and final deuteron are restricted by
\begin{equation}\label{approximation}
|\vec p\,|\sim |\vec p\,'|\ll Q.
\end{equation}
This means that in all expansions one has to keep terms linear in $\vec p$ and $\vec p\,'$ only and
\bee
&E_{1,2}\approx {\textstyle\frac12}E_d \pm \ds{\frac{(\vec d\cdot\vec p\,)}{E_d}}, \quad
E\,'_{1,2}\approx {\textstyle\frac12}E_d \pm \ds{\frac{(\vec d\,'\cdot\vec p\,')}{E_d},\label{E_1}}\\
&\vec p^{\,(1,2)}=-\frac14\vec q \pm \vec p, \quad \vec p\,'^{(1,2)}=\frac14\vec q \pm \vec p\,',\label{p_1}\\
&\Delta_{1,2}=\tfrac12 q\pm \delta, \quad\delta=\left(-\ds{\frac{Q}{2E_d}}(p'_3+p_3),\,\vec p\,'-\vec p\right).
\label{Delta1}
\eee
One sees that in the framework of our approximation all nucleons become on-shell and one can use an expansion similar to (\ref{Expansion1}) for $p\!\!\!/'^{(1)}+m$ and $p\!\!\!/^{(2)}-m$. As a result all 
diagrams of Fig.~\ref{fig:Poles} give the same contribution and
\bee
&&dT^{\mu\nu}_\mathrm{P}\approx -\frac{e^2d^3pd^3p'}{(2\pi)^62E_12E_2'}\times\nonumber\\
&&\times\sum_{\sigma_1,\sigma_2,\sigma'_1,\sigma'_2} 
\langle\vec p\,'^{(2)}\sigma'_2;c|\phi^{\lambda'}|\vec p\,'^{(1)}\sigma'_1\rangle \times\label{Tmunu}\\
&&\times\langle\vec p\,'^{(1)}\sigma'_1|\Gamma^\mu_1(\Delta_1^2) |\vec p\,^{(1)}\sigma_1\rangle \langle\vec p\,^{(1)}\sigma_1|
\phi^{\lambda}|\vec p\,^{(2)}\sigma_2;c\rangle\times\nonumber\\
&&\times
\langle\vec p\,^{(2)}\sigma_2;c|
\bar\Gamma^\mu_2(\Delta_2^2)|\vec p\,'^{(2)}\sigma'_2;c\rangle. \nonumber
\eee
$dT^{\mu\nu}_\mathrm{X}$ is obtained by the exchange $\Delta_1^2 \leftrightarrow\Delta_2^2$ in (\ref{Tmunu}). Using the fact that $\langle\vec p\,^{(1)}\sigma_1|\phi^{\lambda}|\vec p\,^{(2)}\sigma_2;c\rangle$ and $\langle\vec p\,'^{(2)}\sigma'_2;c|\phi^{\lambda'}|\vec p\,'^{(1)}\sigma'_1\rangle $ are Lorentz invariants, one can substitute the non-relativistic deuteron wave functions (\ref{WaveFunction}) instead of these wave functions:
\be
\begin{split}
&\frac{1}{\sqrt{E_d}}\langle\vec p\,^{(1)}\sigma_1|\phi^{\lambda}|\vec p\,^{(2)}\sigma_2;c\rangle
\to(2\pi)^{3/2}\Psi_{\sigma_1\sigma_2}(\lambda,\vec{\widetilde p}\;),\\
&\frac{1}{\sqrt{E_d}}\langle\vec p\,'^{(2)}\sigma'_2;c |\phi^{\lambda'}|\vec p\,'^{(1)}\sigma'_1\rangle
\to(2\pi)^{3/2}\Psi^{\dag}_{\sigma'_1\sigma'_2}(\lambda,\vec{\widetilde p}\;').
\end{split}
\label{NRDWF}
\ee
This substitution must be completed by the transformation of the current $\bar\Gamma^\nu_2\to \Gamma^\nu_2$. In (\ref{NRDWF})
\begin{equation}
\label{argument} 
\begin{split}
&\vec{\widetilde p}=\left( p_1,p_2,{\textstyle\frac{M}{E_d}p_3}\right)=\left(\vec p_\perp,{\textstyle\frac{M}{E_d}p_3}\right),
\\
&\vec{\widetilde p}\;'=\left( {p_1}',{p_2}',{\textstyle\frac{M}{E_d}{p_3}'}\right)=\left(\vec p\;'_\perp,{\textstyle\frac{M}{E_d}{p_3}'}\right)\end{split}
\end{equation}
are the internal momenta in the deuteron rest frame.

Expanding the current matrix elements $\langle \vec p{\;'^{(i)}}\,\sigma'|\Gamma^\mu_i(\Delta_i^2)| \vec p{\;^{(i)}}\,\sigma\rangle=2m\chi^\dag_{\sigma'}\widetilde\Gamma^\mu_i(\Delta_i^2)\chi_\sigma$ in terms of Pauli spinors one gets
\be\label{M_II.interm}
\begin{split}
&\mathcal M^{\mathrm{II}}=
-\frac{64\alpha^2(4\pi)^2 E_d}{ Q^6}\int \dfrac{d^3\widetilde p\,d^3\widetilde p\,'}{(2\pi)^3}\times\\
&\times
\Psi^\dag(\lambda',\vec{\widetilde p}\,')\left[ \tau^{\mu\nu}_h\widetilde \Gamma_{1\mu}(\tfrac14 Q^2)\widetilde \Gamma_{2\nu}(\tfrac14 Q^2)+\vec{\widetilde p}\vec A+\right.\\
&\left.+\vec{\widetilde p}\,'\vec B +\mathcal O(\widetilde p^{\,2},\widetilde p\,'^2)\right]
\Psi(\lambda,\vec{\widetilde p}\,),
\end{split}
\ee
where $\vec A$ and $\vec B$ are some vectors and
\be
\begin{split}
\tau^{\mu\nu}_h&=\bar u_{h}(k')\left[\gamma^\mu(k\!\!\!/-\tfrac12 q \!\!\!/)\gamma^\nu + \gamma^\nu(k\!\!\!/' +\tfrac12 q \!\!\!/)\gamma^\mu\right]u_h(k)=\\
&=j^\nu(k+k')^\mu +  j^\mu(k+k')^\nu.
\end{split}
\label{TPE.2.b}
\ee
The integrals $\int d^3\widetilde p \;\vec {\widetilde p}\Psi(\lambda,\vec{\widetilde p}\,)$ and  $\int d^3\widetilde p\,'\, \vec {\widetilde p}\,'\Psi(\lambda',\vec{\widetilde p}\,'\,)$ obviously vanish after angular integration and one arrives at
\be\label{rough_estimation}
\mathcal M^{\mathrm{II}}\approx 
-\frac{64\alpha^2(4\pi)^2 E_d}{Q^6}\,\tau^{\mu\nu}_h\mathfrak{M}_{\mu\nu}^{\lambda'\lambda},
\ee
where $\mathfrak{M}_{\mu\nu}^{\lambda'\lambda}=\psi_{\lambda'}^\ast(0)\widetilde\Gamma_{1\mu}(\tfrac14 Q^2)\widetilde \Gamma_{2\nu}(\tfrac14 Q^2)\psi_\lambda(0)$; $\psi_{\lambda'}(0)$, $\psi_\lambda(0)$ and $\psi_{\lambda'}(0)$ are the deuteron wave functions in coordinate space at $\vec r=0$ and
\be\label{N_current}
\begin{split}
&\widetilde \Gamma_k^0\left({\textstyle\frac14}Q^2\right)=G_E^k\left({\textstyle\frac14}Q^2\right),\\
&\vec{\widetilde \Gamma}_k\left({\textstyle\frac14}Q^2\right)=\frac i{2M}(\vec\sigma \times \vec q\,)G_M^k\left({\textstyle\frac14}Q^2\right)
\end{split}
\ee
(suffix $k=1,2$ enumerates the nucleons).

For further calculations it is useful to introduce ``plus'' and ``minus'' components of the tensors according to $A_{\pm}=\sqrt{\frac12}(A_1\pm i A_2)$. The contraction of the lepton and deuteron tensors becomes
\begin{equation}
\label{Contracting1}
\begin{split}
\tau^{\mu\nu}\mathfrak{M}_{\mu\nu}^{\lambda'\lambda}=\mathfrak{M}_{00}^{\lambda'\lambda}\tau_{00}-2\left(\mathfrak{M}_{0+}^{\lambda'\lambda}\tau_{0-}+\mathfrak{M}_{0-}^{\lambda'\lambda}\tau_{0+}\right)+\\
+\left(\mathfrak{M}_{++}^{\lambda'\lambda}\tau_{--}+2\mathfrak{M}_{+-}^{\lambda'\lambda}\tau_{-+}+\mathfrak{M}_{--}^{\lambda'\lambda}\tau_{++}\right),
\end{split}
\end{equation}
where
\begin{equation}
\label{LeptonComps}
\begin{split}
&\tau_{00}=8E_e^2\cos\textstyle{\frac{\theta}2},\\
&\tau_{0+}=-2\sqrt{2} E_e^2\left(2- \sin^2\textstyle{\frac{\theta}2}-h\sin\textstyle{\frac{\theta}2}\right),\\
&\tau_{0-}=-2\sqrt{2} E_e^2\left(2-\sin^2\textstyle{\frac{\theta}2}+h\sin\textstyle{\frac{\theta}2}\right),\\
&\tau_{++}=4 E_e^2\cos\textstyle{\frac{\theta}2}\left(1-h\sin\textstyle{\frac{\theta}2}\right),\\
&\tau_{--}=4 E_e^2 \cos\textstyle{\frac{\theta}2}\left(1+h\sin\textstyle{\frac{\theta}2}\right),\\
&\tau_{-+}=4 E_e^2\cos\textstyle{\frac{\theta}2}
\end{split}
\end{equation}
and
\be
\label{M0_ll_S-wave}
\begin{split}
&\mathfrak{M}_{00}^{11}=\mathfrak{M}_{00}^{-1-1} = \mathfrak{M}_{00}^{00}=\frac{\mathcal{C}}{4\pi}G_{EE},\\
&\mathfrak{M}_{0+}^{10}=\mathfrak{M}_{0+}^{0-1}=-\mathfrak{M}_{0-}^{01} =-\mathfrak{M}_{0-}^{-10}=\\
&\hspace{.75cm}=-\frac{\mathcal{C}}{4\pi }\sqrt{\eta}\, G_{EM},\\
&\mathfrak{M}_{++}^{1-1}=\mathfrak{M}_{--}^{-11} = \frac{\mathcal{C}}{2\pi}\eta \,G_{MM},\\
&\mathfrak{M}_{\pm\mp}^{00} = -\frac{\mathcal{C}}{4\pi}\eta \,G_{MM},
\end{split}
\ee
with the abbreviations
\begin{eqnarray*}
&\mathcal C=\left.\left[u'_0(r)\right]^2\right|_{r=0} ,\\
&G_{EE}=G_E^p(\tfrac14Q^2)G_E^n(\tfrac14Q^2),\\
&G_{MM}=G_M^p(\tfrac14Q^2)G_M^n(\tfrac14Q^2),\\
&G_{EM}=\tfrac12\left[G_E^p(\tfrac14Q^2)G_M^n(\tfrac14Q^2) +G_M^p(\tfrac14Q^2)G_E^n(\tfrac14Q^2)\right].
\end{eqnarray*}
Finally we get the following amplitudes

\be 
\label{T_ll_II_ampl}
\begin{split}
&T^\mathrm{II}_{11}=
\varkappa\cos\tfrac{\theta}2G_{EE},
\\
&T^\mathrm{II}_{00}=
\varkappa\cos \tfrac{\theta}2\left(G_{EE} -\eta G_{MM}\right),
\\
&T^\mathrm{II}_{10,h}=-\varkappa\sqrt{\tfrac{\eta}{2}}G_{EM}\left(2-\sin^2\tfrac{\theta}2+h\sin\tfrac{\theta}2\right),
\\
&T^\mathrm{II}_{1-1,h}=\varkappa\eta G_{MM}\cos\tfrac{\theta}2 \left(1+h\sin\tfrac{\theta}2\right),
\end{split}
\ee
where
\be\label{kappa}
\varkappa=-\dfrac{128\alpha\mathcal C E_e}{Q^4}
\ee
and one arrives at
\be
\label{FF_II}
\begin{split}
&\delta\mathcal G_C^\mathrm{II}=\varkappa\left(G_{EE} -\tfrac13\eta G_{MM}\right),\\
&\delta\mathcal G_Q^\mathrm{II}=-\dfrac{\varkappa}{2}G_{MM},\\
&\delta\mathcal G_M^\mathrm{II}=\dfrac{2\varkappa G_{EM}}{1+\sin^2\tfrac{\theta}2},\\
&g_1^\mathrm{II}=\dfrac{\varkappa G_{EM}\cos^2\tfrac{\theta}2}{1+\sin^2\tfrac{\theta}2},\\
&g_2^\mathrm{II}=g_3^\mathrm{II}=\varkappa\eta\cos\tfrac{\theta}2 G_{MM}.
\end{split}
\ee
\begin{figure}
  \centering 
\includegraphics[height=0.25\textheight]{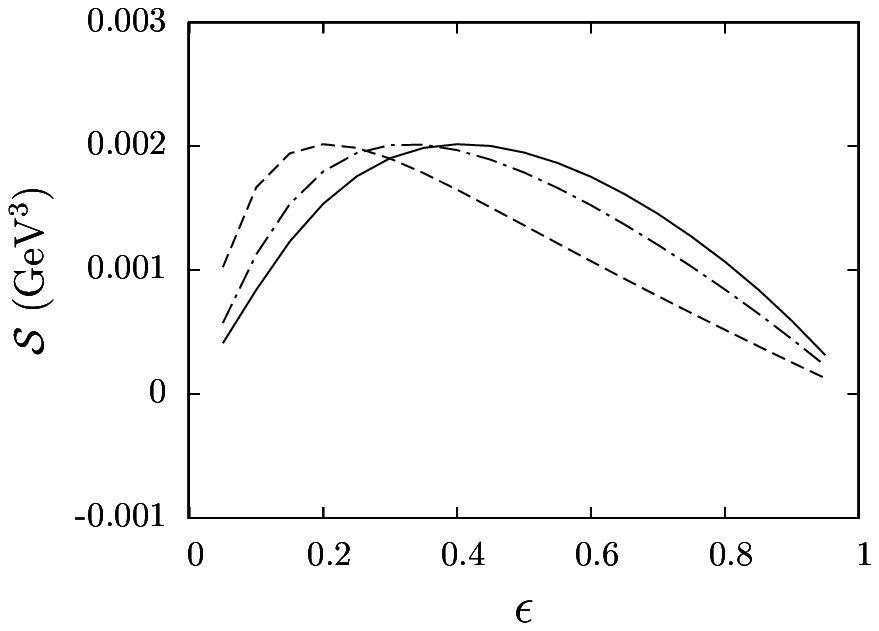}
\includegraphics[height=0.25\textheight]{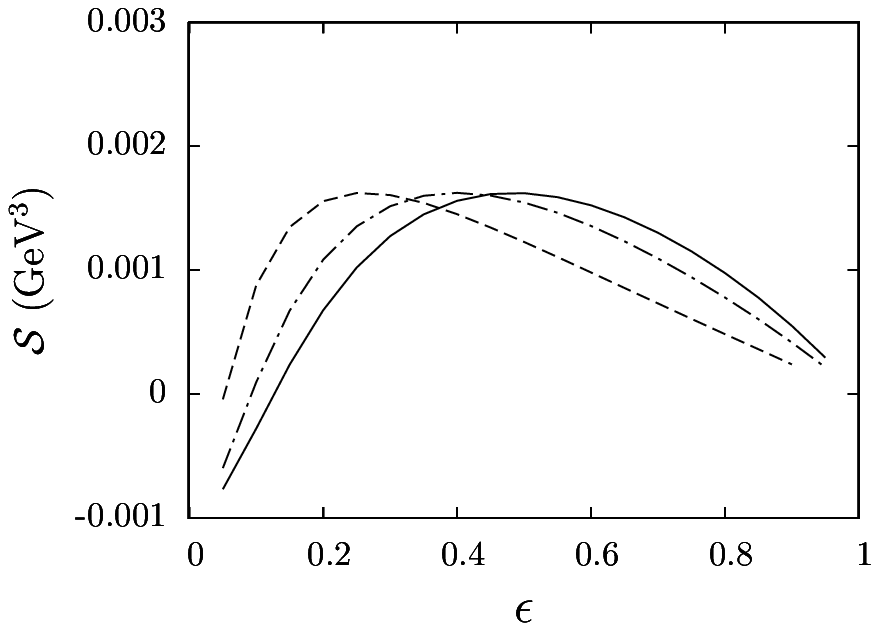}
\caption{\label{fig:StoS} $\mathcal S$ factor calculated for the CD-Bonn potential \cite{CD-Bonn} (left) and the Paris potential \cite{Paris} (right). Dashed, dot-dashed and solid curves are for $Q^2=$1,~2~and~3~GeV$^2$, respectively.}
\end{figure}

One should note that the approximation (\ref{M_II.interm}) is not valid at $\theta\to 0$ (or, equivalently, $\epsilon\to 1$). Indeed, the denominator of the electron propagator 
$$
\frac14 Q^2+\frac{E_e Q}{M}(\widetilde p_z+\widetilde p\,'_z)-2\cos\tfrac{\theta}2E_e(\widetilde p_x-\widetilde p\,'_x)Q^2
$$
contains products of $E_e$ and components of the internal momenta. From (\ref{Breitframe0}) it follows that $E_e\to \infty$ when $\quad \theta \to 0$ and the factor $\mathcal C$ should be changed to
\be\label{S_prime}
\begin{split}
&\mathcal C \to \mathcal S=\\
&=\frac{1}{(2\pi)^3}\int\frac{d^3\widetilde pd^3\widetilde p\,' U_0(\widetilde p)U_0(\widetilde p\,')}
{1+\frac{4E_e}{QM}(\widetilde p_z+\widetilde p\,'_z)-8\cos\tfrac{\theta}2\frac{E_e(\widetilde p_x-\widetilde p\,'_x)}{Q^2} +i0}.
\end{split}
\ee
The amplitudes (\ref{T_ll_II_ampl}) with the substitution (\ref{S_prime}) in (\ref{kappa}) coincide with the results of Ref.~\cite{Lev}.

To evaluate the integral above one can use the integral representation for the denominator
\be\label{IR}
\dfrac{1}{\alpha +i0}=-i\int_0^\infty d\tau e^{i(\alpha +i0)\tau}
\ee
and reduce (\ref{S_prime}) to a one-dimensional integral
\be\label{S_prime_1}
\mathcal S=-\tfrac{i}4 Q^2\int_0^\infty \dfrac{d\tau}{y^2}e^{\frac i4 Q^2\tau}u^2_0(y),
\ee
where $y=\tau E_e\sqrt{4\cos^2\tfrac{\theta}2 +\frac{Q^2}{M^2}}$. Changing the variable in (\ref{S_prime_1}) one gets
\be\label{S_prime2}
\mathcal S=-if\int_0^\infty \dfrac{dy}{y^2}e^{ify} u^2_0(y),
\ee
where 
\be\label{f_real}
f=\dfrac{Q^2}{4E_e\sqrt{4\cos^2\tfrac{\theta}2 +\frac{Q^2}{M^2}}}.
\ee
For the standard parametrization of the wave function
\be\label{WF_0}
u_0(y)=\sum_nc_n e^{-\alpha_n y},\qquad\text{with}\qquad  \sum_nc_n=0
\ee
we obtain (Appendix~\ref{app:integral})
\be\label{S_prime3}
\mathcal S=-if\sum_n\sum_mc_nc_m(\alpha_n+\alpha_m-if)\ln(\alpha_n+\alpha_m-if).
\ee
In (\ref{S_prime2}) the exponent reduces to 1 in the limit $E_e\to \infty$ for fixed 
$Q$ and
\be\label{S_prime_asy}
\Re\mathrm e T^{\mathrm{II}}_{\lambda'\lambda}\sim \sin\tfrac{\theta}2,\quad
\Im\mathrm m T^{\mathrm{II}}_{\lambda'\lambda}\to const,
\ee
i.e., at the limit $\theta \to 0$ the TPE does not contribute to the cross-section in the next order of the $\alpha$-expansion.

The $\epsilon$ and $Q^2$ dependence of $\mathcal S$ is displayed in Fig.~\ref{fig:StoS}. One sees that the $\mathcal S$ factor depends strongly on the NN potential and in
any case it is very different from the constant value $\mathcal C=\left[ u_0'(0)\right]^2 $. The reason is as follows: from (\ref{S_prime3}) one gets that $\mathcal S \to \mathcal C$ at the formal limit 
\be\label{f_limit}
f\gg \alpha_n+\alpha_m.
\ee
But from expression (\ref{f_real}) it follows that $f\to \frac12 M\sin\frac{\theta}2$ 
when $Q^2 \to \infty$ and the condition  (\ref{f_limit}) cannot be fulfilled at 
any $Q^2$. Note that a similar situation takes place in the evaluation of the 
so-called triangle diagram in $pd$ backward scattering \cite{KolybasovSmorodinskaya}.

We have also studied the deuteron $D$-wave contribution to the $\mathcal S$ factor and 
found that it contributes less than 10\%.
\begin{table}
\caption{$u'_0(0)$ for some popular potentials.
\label{table1}}
\begin{ruledtabular}
\begin{tabular}{ccc}
$u'(0)$, fm$^{-3/2}$ & Potential& Ref.\\ 
\tableline
$1.1978\cdot 10^{-1}$ & Paris   & \cite{Paris}\\
$3.1035\cdot 10^{-1}$ & CD-Bonn & \cite{CD-Bonn}\\ 
$2.6860\cdot 10^{-1}$ & Nijm I  & \cite{NIJM}\\ 
$2.6730\cdot 10^{-2}$ & Nijm II & \cite{NIJM}\\ 
$3.1571\cdot 10^{-1}$ & Nijm 93 & \cite{NIJM}\\
$5.8334\cdot 10^{-2}$ & Reid 93 & \cite{NIJM}\\
\end{tabular}
\end{ruledtabular}
\end{table}
\begin{figure*}
\includegraphics[height=0.25\textheight]{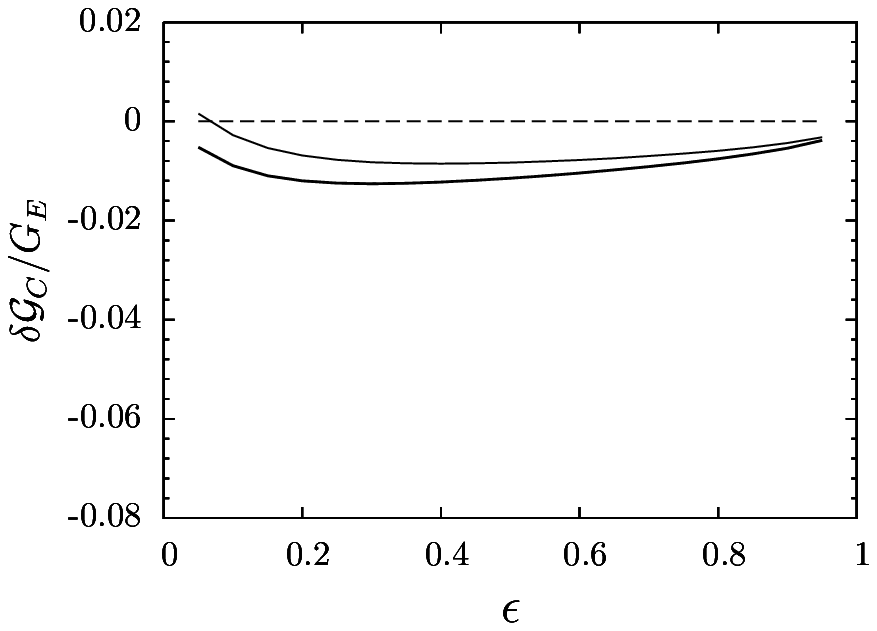}
\includegraphics[height=0.25\textheight]{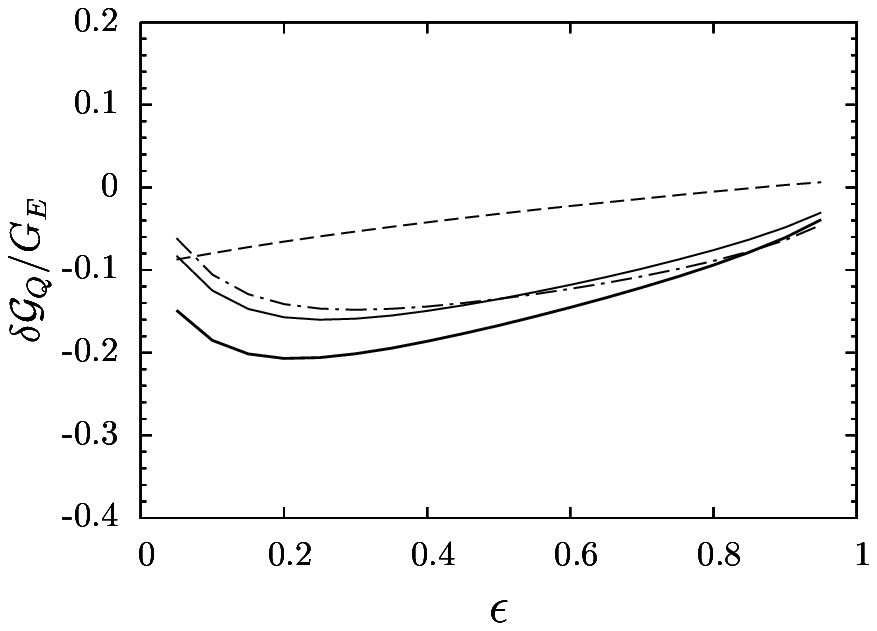}\\ 
\includegraphics[height=0.25\textheight]{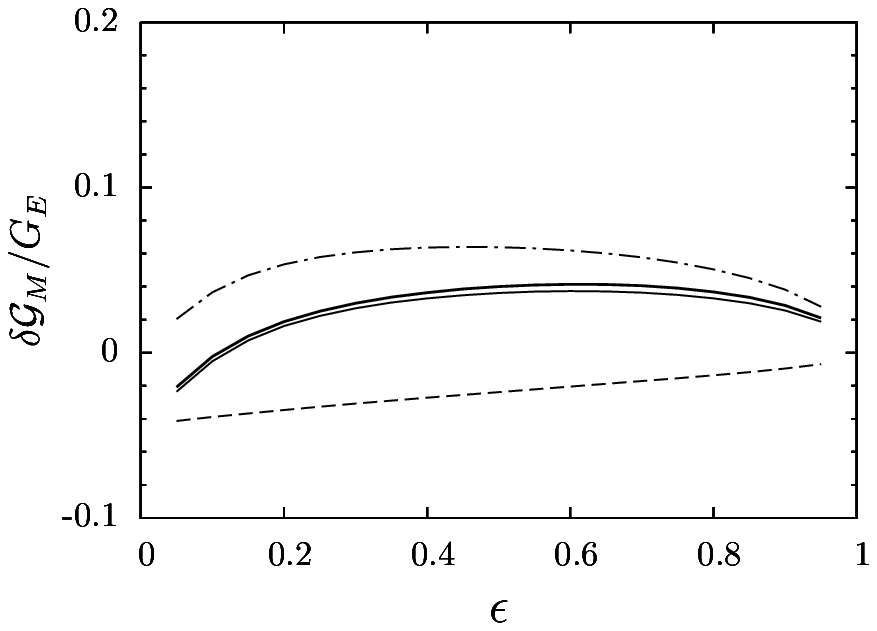}
\includegraphics[height=0.25\textheight]{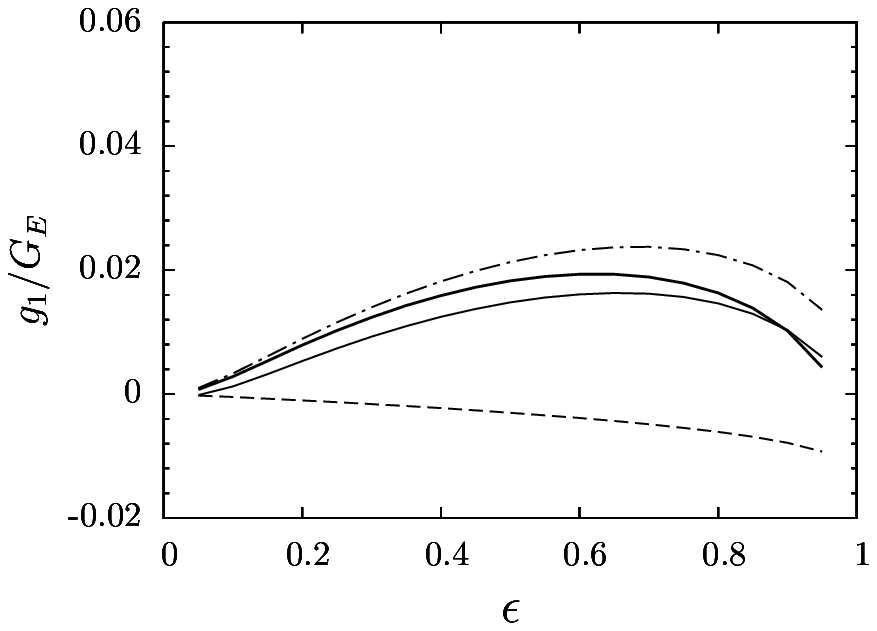}
\caption{\label{fig:Q^2=1}Two-photon exchange corrections $\delta \mathcal G_C/G_E$, $\delta \mathcal G_Q/G_E$,  $\delta \mathcal G_M/G_E$ and  $g_1/G_E$ at $Q^2=$1~GeV$^2$. Dashed, dot-dashed and solid (bold) curves are for $\mathcal M^{\mathrm I}$, $\mathcal M^{\mathrm{II}}$ and $\mathcal M^{\mathrm I}+\mathcal M^{\mathrm{II}}$, respectively, calculated with the CD-Bonn potential. These solid (thin) curves depict $\mathcal M^{\mathrm I}+\mathcal M^{\mathrm{II}}$ calculated with the Paris potential.}
\includegraphics[height=0.25\textheight]{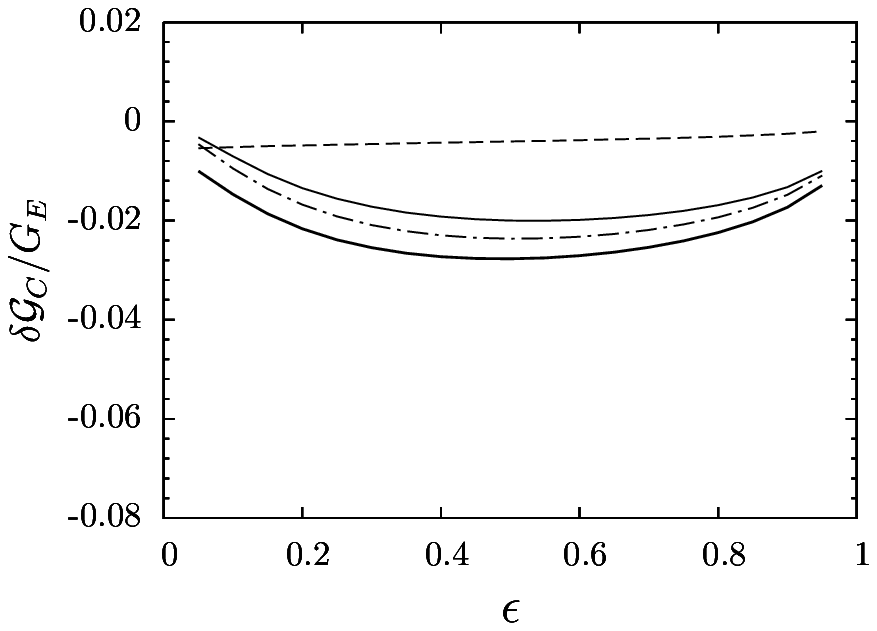}
\includegraphics[height=0.25\textheight]{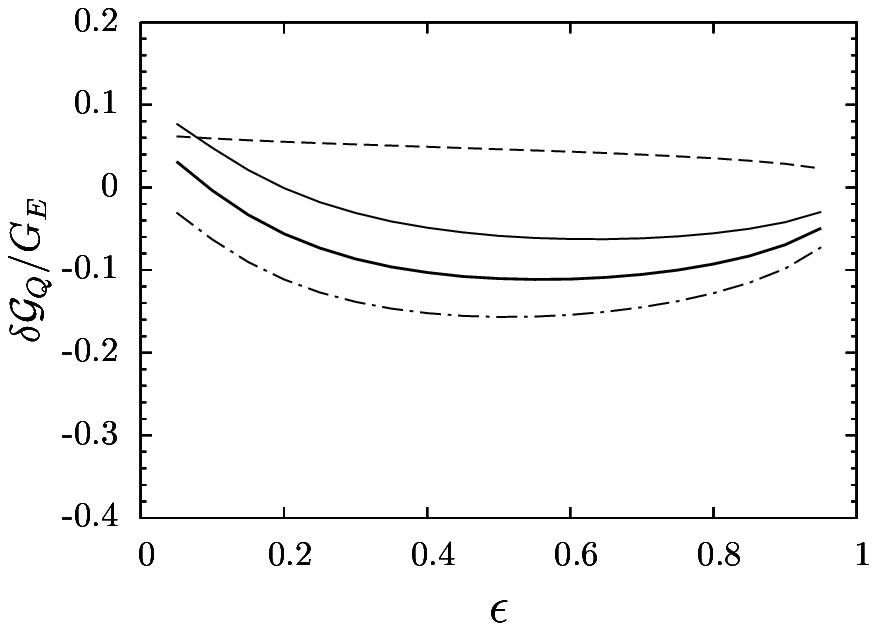}\\ 
\includegraphics[height=0.25\textheight]{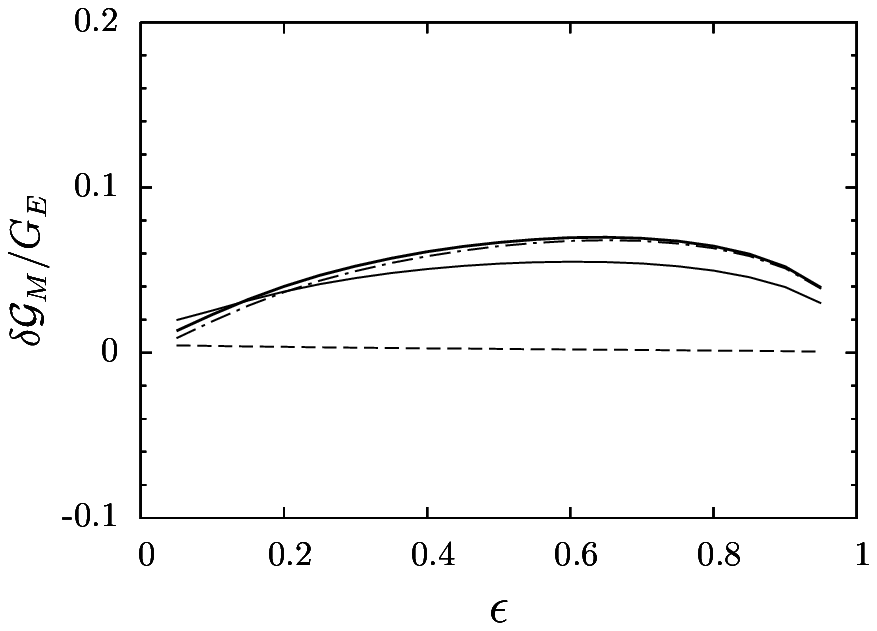}
\includegraphics[height=0.25\textheight]{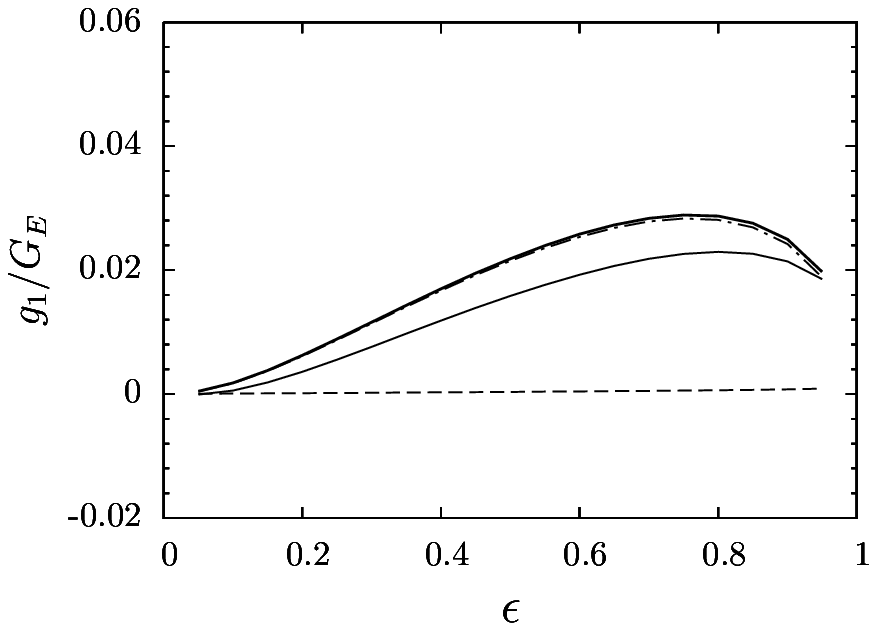}
\caption{\label{fig:Q^2=2} Same as Fig.~\ref{fig:Q^2=1} for $Q^2=$2~GeV$^2$.}
\end{figure*}%
\begin{figure*}
\includegraphics[height=0.25\textheight]{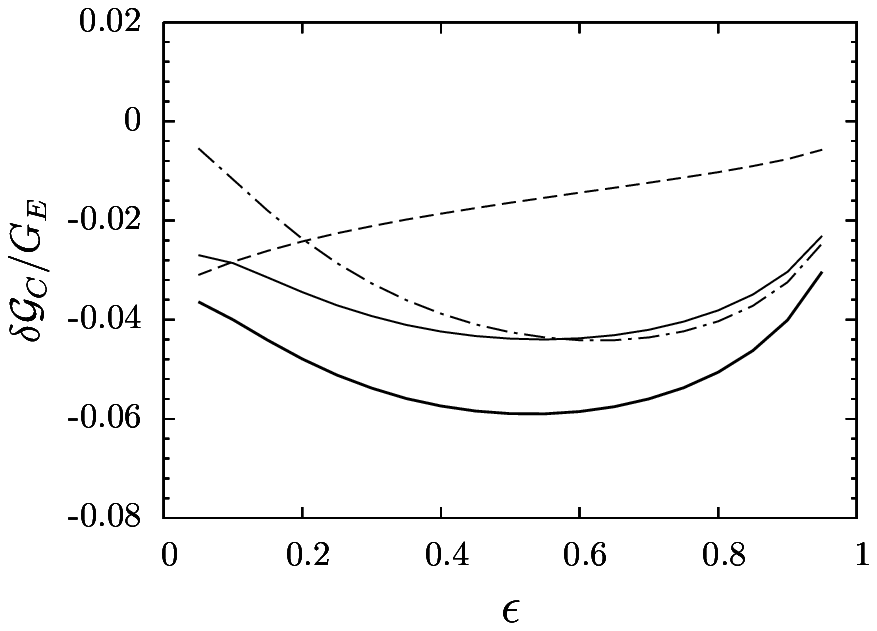}
\includegraphics[height=0.25\textheight]{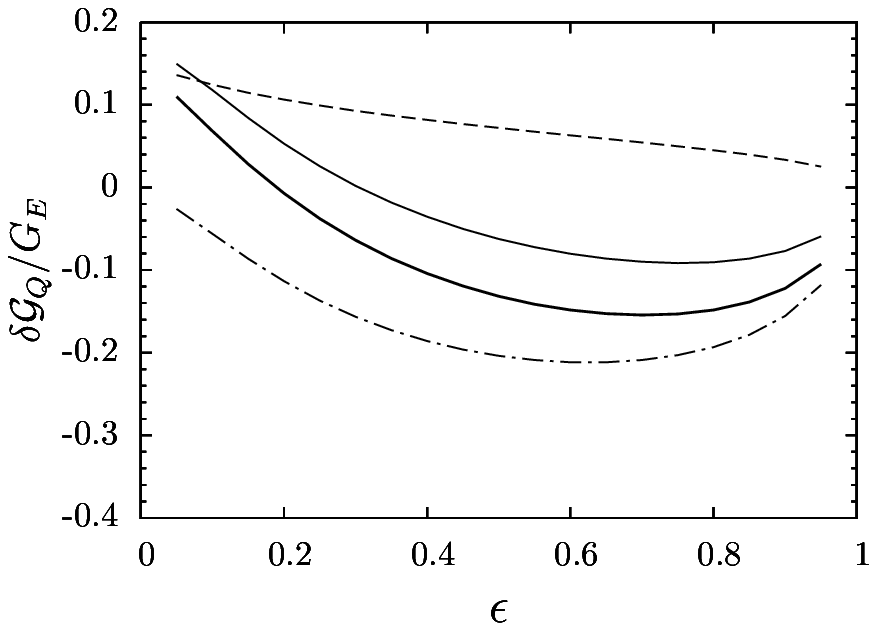}\\ 
\includegraphics[height=0.25\textheight]{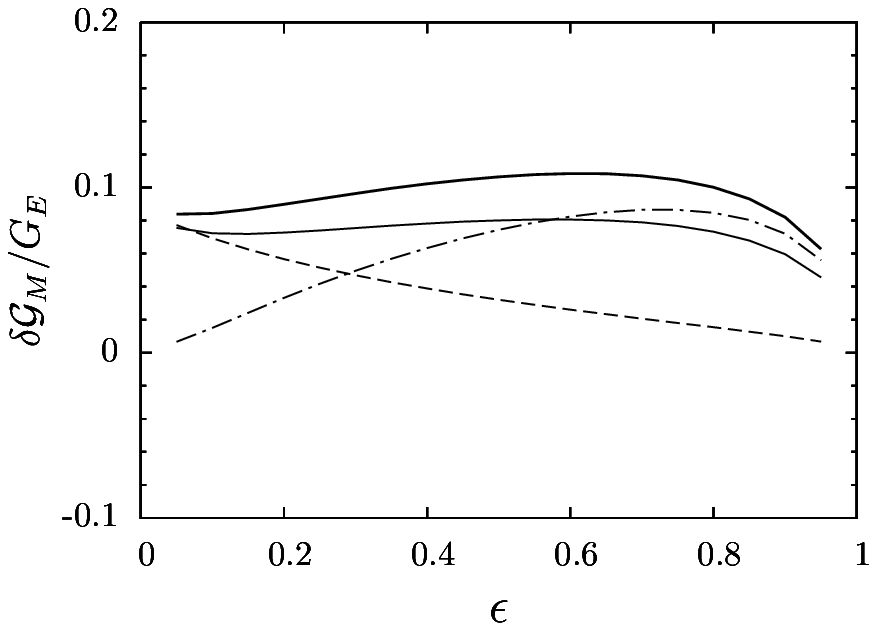}
\includegraphics[height=0.25\textheight]{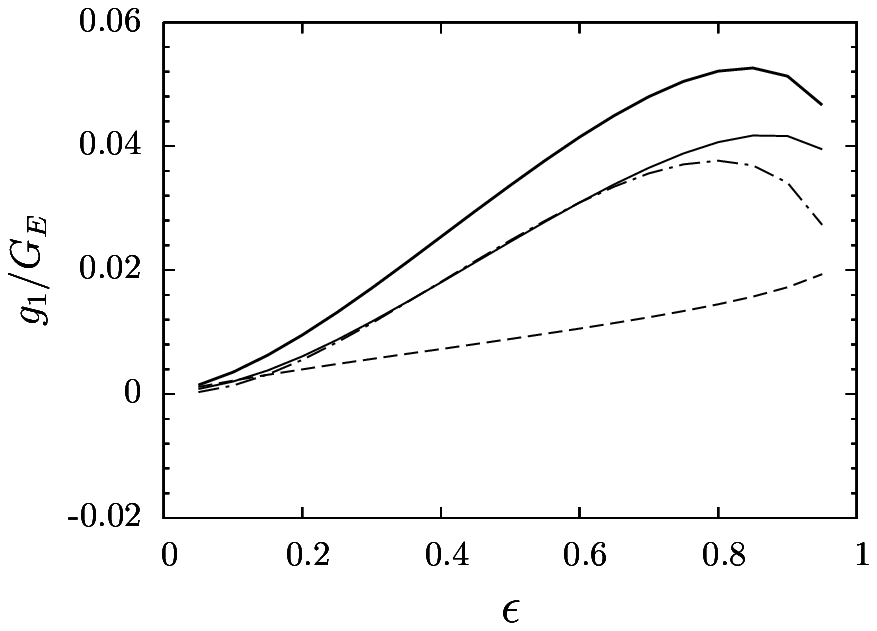}
\caption{\label{fig:Q^2=3} Same as Fig.~\ref{fig:Q^2=1} for $Q^2=$3~GeV$^2$.}
\includegraphics[height=0.25\textheight]{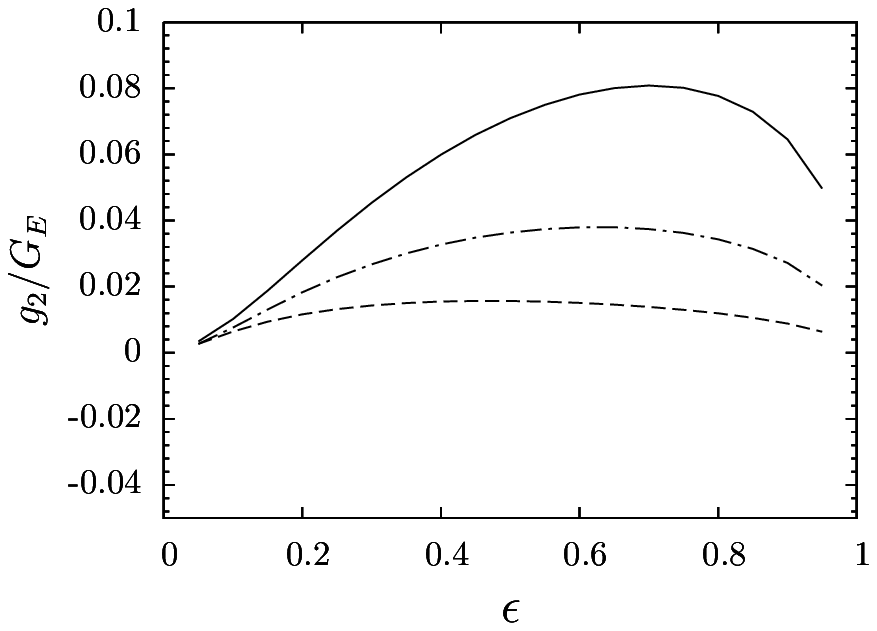}
\includegraphics[height=0.25\textheight]{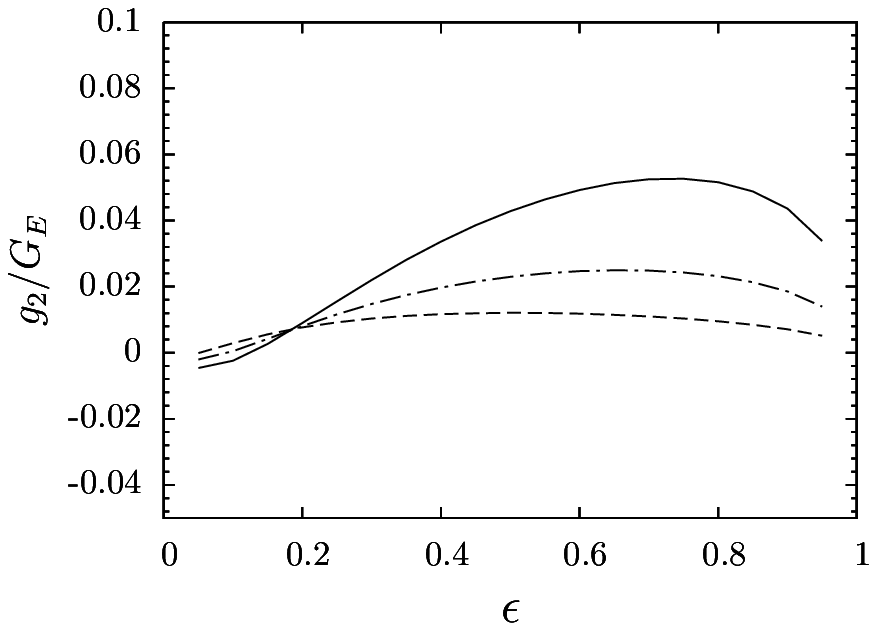}
\caption{\label{fig:g23} Ratio of $g_2=g_3$ to $G_E$. Dashed, dot-dashed and solid curves are for $Q^2=$1,~2~and~3~GeV$^2$, respectively. Left and right panels are for the CD-Bonn and Paris potentials, respectively.}
\end{figure*}

\section{Numerical results and discussion\label{sec:num}}
Figs.~\ref{fig:Q^2=1}-\ref{fig:g23} display the $\epsilon$-behavior of the TPE 
corrections $\delta\mathcal G_C/G_E$, $\delta\mathcal G_Q/G_E$, $\delta\mathcal G_M/G_E$ 
and $g_{1,2}/G_E$ calculated with the deuteron wave function for the CD-Bonn and 
the Paris potentials. The form factor $G_E(Q^2)$ was calculated in the framework of the 
impulse approximation.

In the present calculations of the TPE correction in $\mathcal M^{\mathrm{I}}$ the 
amplitudes $\Delta \widetilde F^N_{1,2}$ and $\widetilde F_3$ from the theoretical 
calculations of Ref.~\cite{BorisyukKob} are used. At $Q^2<6\ \mathrm{GeV}^2$ they are practically independent of the parametrization of the nucleon form factor. For $\mathcal M^{\mathrm{II}}$ we use the following parametrization of the nucleon form factor: 
\begin{itemize}
 \item for the magnetic form factors of the proton and neutron --- dipole parametrization
\be \label{FFN_param_mag}
G_M^p(Q^2)=\mu_pG_D(Q^2), \qquad G_M^n(Q^2)=\mu_nG_D(Q^2),
\ee
where $G_D(Q^2)=\left(1+{Q^2}/{0.71}\right)^{-2}$;
\item the electric form factors of the proton and neutron were taken from the parametrization of the JLab data (see \cite{PPV})
\be \label{JLab_param}
G_E^p(Q^2)=\left(1.0587-0.14265Q^2\right)G_D(Q^2),
\ee
and and so-called Galster parametrization \cite{Gastler}, respectively.
\end{itemize}

One sees that two-photon exchange may give a large contribution to the 
elastic $ed$ scattering; although caution is required as these estimates have large 
uncertainties. The most important source of uncertainty comes from the 
$\mathcal S$ factor which is determined by the short range part of the deuteron 
wave function. The last quantity is very poorly known (see, e.g., 
Table~\ref{table1}). 
Of course, besides NN degrees of freedom, non-nucleon (quark) degrees of freedom 
should also be taken into account in this region and one may expect that in the 
framework of more realistic estimates the two-photon corrections may be smaller. 
The implication is that the experimental study of two-photon exchange in elastic $ed$ 
scattering at $Q^2\sim$\,few~GeV$^2$ can give important information about the 
deuteron structure at short distances.

In summary, we estimated the two-photon exchange amplitude in elastic $ed$ scattering. 
There are six independent form factors which determine this amplitude, but only three 
of them contribute to the cross section in second order perturbation theory.

There are two types of two-photon exchange diagrams. For the first type two intermediate photons interact with the same nucleon. For the second type the intermediate photons interact with different nucleons.

We show that the two-photon exchange amplitude is strongly connected with the deuteron structure at short distances.
\begin{acknowledgements}
The authors thank A.-Z.~Dubni\v ckova for important discussions and D.L.~Borisyuk for providing the numerical calculations of two-photon exchange form factors for the electron-nucleon scattering. We are also grateful to D.L.~Borisyuk, M.~Faber and C.F.~Perdrisat for reading the manuscript and critical remarks. This work was partially supported by a Joint Research Project between the Ukrainian and Slovak Academies of Sciences and the Slovak Grant Agency for Sciences VEGA, grant No.2/0009/10.
\end{acknowledgements}
\appendix
\section{\label{AppendixOPE}General expression for the $ed\to ed$ amplitude in the Breit frame}

From the invariance under Lorentz transformations and space and time inversions it follows that the amplitude of the elastic scattering of a spin-$\frac12$ particle (the electron) off a spin-1 particle (the deuteron) has 9 invariant amplitudes (form factors). Usually for ultra-relativistic electrons the mass can be neglected and the electron helicity is conserved. In this case the number of form factors is reduced to six and the amplitude has the general form \cite{GakhTomasi,DongPRC}
\begin{equation}
\label{OPE.1.A}
\mathcal M=-\frac{4\pi \alpha}{q^2}j^\mu J_\mu\equiv \frac{4\pi \alpha}{Q^2}T_{\lambda'\lambda,h},
\end{equation} 
where $j_\mu$ is the e.m. current for the electron and $J^\mu$ is an ``effective current'' for the deuteron
\be\label{deuteronEMcurrent}
\begin{split}
&J_\mu=\\&=-\left\{
G_1\left( {\epsilon'}^\ast\cdot\epsilon\right)(d+d')_\mu + \right.\\
&\hspace{0.6cm} 
+
G_2\left[\left( {\epsilon'}^\ast\cdot q\right)\epsilon_\mu-{\epsilon'_\mu}^\ast \left( \epsilon\cdot q\right) \right]- \\
&\hspace{0.6cm}
-G_3\frac{\left({\epsilon'}^\ast\cdot q\right)\left(\epsilon\cdot q\right)}{2M^2}(d+d')_\mu+\\
&\hspace{0.6cm} +G_4\dfrac{\left( {\epsilon'}^\ast\cdot K\right)\left(\epsilon\cdot K\right)}{2M^2}(d+d')_\mu+\\
&\hspace{0.6cm} 
+G_5 \left[\left( {\epsilon'}^\ast\cdot K\right)\epsilon_\mu+{\epsilon'_\mu}^\ast\left(\epsilon\cdot K\right)
\right] +\\
&\left.\hspace{0.6cm} 
+G_6\dfrac{
\left( {\epsilon'}^\ast\cdot K\right) \left(\epsilon\cdot q\right)-\left({\epsilon'}^\ast\cdot q\right)\left(\epsilon\cdot K\right)}{2M^2}(d+d')_\mu
\right\}.
\end{split}
\ee
Here $K=k+k'$. The form factors $G_1,...,G_6$ are complex functions of two variables, e.g., $Q^2$ and $\theta$.

In the Breit frame one simply finds
\begin{equation}
\label{OPE.5}
\begin{split}
&\epsilon_{(\pm)}\cdot q=\epsilon'_{(\pm)}\cdot q=0, \\ 
&\epsilon_{(0)}\cdot q=\epsilon'_{(0)}\cdot q=-\frac{E_dQ}M,\\
&{\epsilon'}^\ast_{(\pm)}\cdot \epsilon_{(\pm)}=-1, \quad {\epsilon'}^\ast_{(\mp)}\cdot \epsilon_{(\pm)}=0,\\
&\epsilon_{(\pm)}\cdot \epsilon_{(0)}=\epsilon_{(\pm)}\cdot \epsilon'_{(0)}=0, \\ 
&{\epsilon'}^\ast_{(0)}\cdot \epsilon_{(0)}=-\left(1+\frac{Q^2}{2M^2}\right),\\
&\epsilon_{(\pm)}\cdot K=\epsilon'_{(\pm)}\cdot K=\pm \sqrt{2}E_e\cos\tfrac{\theta}2,\\
&\epsilon_{(0)}\cdot K=-\epsilon'_{(0)}\cdot K=-\frac{E_eQ}{M}
\end{split}
\end{equation}
and
\begin{equation}
\label{OPE.6}
\begin{split}
&\epsilon_{(0)}\cdot j=-\frac{E_e Q}{M}\cos\tfrac{\theta}2, \quad
\epsilon'_{(0)}\cdot j=\frac{E_e Q}{M}\cos\tfrac{\theta}2,\\
&\epsilon_{(\pm)}\cdot j=\sqrt2E_e(\pm1-h\sin\tfrac{\theta}2), \\
&\epsilon'_{(\pm)}\cdot j=\sqrt2E_e(\pm1+h\sin\tfrac{\theta}2),\\
&(d+d')\cdot j=4E_e E_d\cos\tfrac{\theta}2.
\end{split}
\end{equation} 
From Eqs.~(\ref{OPE.1.A})-(\ref{OPE.6}) it follows that 
\be\label{Gi_ampl}
\begin{split}
&T_{11,h}=T_{-1-1,h}=\\
&=\left[ G_1-\left( \dfrac{E_e}{M}\right) ^2\cos^2\frac{\theta}2 G_4 -\dfrac{E_e}{E_d} G_5\right]\cos\tfrac{\theta}{2} ,
\\
&T_{00,h}=
\left[ \left(1+2\eta \right)G_1 -2\eta G_2 +2(1+\eta)\eta G_3 -\right. \\
&\left. \hspace{0.5cm}-2 \left( \dfrac{E_e}{M}\right) ^2\eta G_4-4\dfrac{E_eE_d}{M^2} G_6\right] \cos\tfrac{\theta}{2},\\
&T_{10,h}=-T_{01,-h}=T_{0-1,h}=-T_{-10,-h}=\\
&=\sqrt2 \frac{Q}{4E_d M}\left[ 
-E_d G_2 + 2\frac{E_d E_e^2}{M^2}\cos^2\tfrac{\theta}{2} G_4+
\right.\\
&\hspace{0.5cm}+E_e\left(1+ \cos\tfrac{\theta}{2}\right)G_5 +2\frac{E_d^2 E_e}{M^2}\cos\tfrac{\theta}{2}G_6+\\
&\hspace{0.5cm}\left.+h\sin \tfrac{\theta}{2}\left( -E_d G_2 +E_e G_5\right) \right],\\
&T_{1-1,h}=T_{-11,-h}=\\
&=\frac{E_e}{E_d}\cos\tfrac{\theta}2\left[\frac{E_e E_d}{M^2} \cos\tfrac{\theta}2 G_4 +\left(1+h\sin \tfrac{\theta}{2} \right)G_5 \right] .
\end{split}
\ee
In the Born approximation the form factors $G_4$, $G_5$ and $G_6$ vanish and the form factors $G_1$, $G_2$ and $G_3$ become real functions $G_1^{(0)}$, $G_2^{(0)}$ and $G_3^{(0)}$ of one variable $Q^2$, that is
\be
\begin{split}
&G_1=G_1^{(0)}(Q^2)+\mathcal O(\alpha),\qquad G_2= G_2^{(0)}(Q^2)+\mathcal O(\alpha),\\ 
&G_3= G_3^{(0)}(Q^2)+\mathcal O(\alpha),\qquad G_4\sim G_5\sim G_6\sim \alpha.
\end{split}
\ee
Commonly the charge, magnetic and quadrupole form factors, $G_C(Q^2)$, $G_M(Q^2)$ and $G_Q(Q^2)$ are used instead of the form factors $G_1^{(0)}$, $G_2^{(0)}$ and $G_3^{(0)}$. They are connected by 
\begin{equation}
\label{OPE.3}
\begin{split}
G_1^{(0)}(Q^2)=&G_C(Q^2)-\tfrac23\eta G_Q(Q^2), \\
G_2^{(0)}(Q^2)=&G_M(Q^2),\\
G_3^{(0)}(Q^2)=&\frac1{1+\eta}\left[
-G_C(Q^2)+G_M(Q^2)+\right.\\
&\left.+\left(1+\tfrac23\eta\right) G_Q(Q^2)
\right]
\end{split}
\end{equation}
and Eqs.~(\ref{Gi_ampl}) are reduced to
\be\label{Gi_ampl_0}
\begin{split}
&T^{(0)}_{11,h}=T^{(0)}_{-1-1,h}=\left[G_C(Q^2)-\tfrac23\eta G_Q(Q^2)\right]\cos\tfrac{\theta}{2} ,
\\
&T^{(0)}_{00,h}=
\left[G_C(Q^2)+\tfrac43\eta G_Q(Q^2)\right] \cos\tfrac{\theta}{2},\\
&T^{(0)}_{10,h}=-T^{(0)}_{01,-h}=T^{(0)}_{0-1,h}=-T^{(0)}_{-10,-h}=\\
&=-\sqrt{\frac{\eta}2}G_M(Q^2)\left(1+h\sin \tfrac{\theta}{2} \right) ,\\
&T^{(0)}_{1-1,h}=T^{(0)}_{-11,-h}=0.
\end{split}
\ee
\section{\label{Appendix1.0}Non-relativistic reduction of the effective hadron current \label{App.TPE.I.3}}
In the Breit frame $K=E_e(1,\cos\tfrac{\theta}2,0,0)$ and
\bee
&&\left\langle \vec p{\,'^{(N)}}\sigma'\left| H_{N}^0 \right|\vec p^{\,(N)}\,\sigma\right\rangle 
\approx \chi^\dag_{\sigma'}\left\lbrace
2m\Delta \widetilde F_{1N}+\right.\\
&&\left.\hspace{0.25cm}
+\frac{Q}{2m}\left( -Q+2i\epsilon^{3nm}p^n\sigma^m\right) \Delta \widetilde F_{2N}+\right. \nonumber\\
&&\left.\hspace{0.25cm}
+\frac{E_e}{m}\left[ 2m-\cos\tfrac{\theta}2 (2p^1+iQ\sigma^2)\right] \widetilde F_{3N}
\right\rbrace
\chi_\sigma,\label{app.red.7} \nonumber\\
&&\left\langle \vec p{\,'^{(N)}}\sigma'\left| H_{N}^a \right|\vec p^{\,(N)}\,\sigma\right\rangle \approx
2\chi^\dag_{\sigma'}\left[
(p^a - \right.\\
&&\left.\hspace{0.25cm} -iQ\epsilon^{a 3 n }\sigma^n)\Delta \widetilde F_{1N}
+iQ\epsilon^{3an}\sigma^n\Delta \widetilde F_{2N}
\right]
\chi_\sigma,\label{app.red.10} \nonumber\\
&&\left\langle \vec p{\,'^{(N)}}\sigma'\left| H_{N}^3 \right|\vec p^{\,(N)}\,\sigma\right\rangle \approx\\
&&\hspace{0.25cm}\approx (p^{(N)}+p{'^{(N)}})^3\Delta \widetilde F_{1N}\chi^\dag_{\sigma'}\chi_\sigma.\label{app.red.12} \nonumber
\eee
where $a=1,2$.

Note that up to terms of order $\mathcal O\left( \frac{p^1}{m}\right) \sim \frac{50\text{ MeV/c}}{m}$ the amplitudes $\Delta \widetilde F_{1N}$, $\Delta \widetilde F_{2N}$ and $\widetilde F_{3N}$ are independent on the nucleon momenta. In this approximation the terms proportional to $p_\perp$ will vanish after integration in Eq.~(\ref{TPE.I.3.b}) and Eqs.~(\ref{app.red.7}) and (\ref{app.red.10}) become
\bee
&&\left\langle \vec p{\,'^{(N)}}\sigma'\left| H_{N}^0 \right|\vec p^{\,(N)}\,\sigma\right\rangle 
\approx \chi^\dag_{\sigma'}\left[
2m\Delta \widetilde F_{1N}
-\frac{Q^2}{2m} \Delta \widetilde F_{2N} + \right.\nonumber\\
&&\left. \hspace{0.5cm}
+\frac{E_e}{m}\left( 2m- iQ\cos\tfrac{\theta}2\sigma^2\right) \widetilde F_{3N}
\right]
\chi_\sigma=\nonumber \\
&& \hspace{0.5cm}=\chi^\dag_{\sigma'}\left(
2m\Delta \mathcal G_E- i\frac{E_eQ}{m}\cos\tfrac{\theta}2\sigma^2\widetilde F_{3N}
\right)
\chi_\sigma,\label{app.red.7.a}\\
&&\left\langle \vec p{\,'^{(N)}}\sigma'\left| H_{N}^a \right|\vec p^{\,(N)}\,\sigma\right\rangle \approx\nonumber\\
&& \hspace{0.5cm}\approx 2iQ\epsilon^{3an}(\Delta \widetilde F_{1N} + \Delta \widetilde F_{2N})
\chi^\dag_{\sigma'}\sigma^n\chi_\sigma.\label{app.red.10.a}
\eee
\section{\label{app:integral}}
Let us consider the integral
\be\label{app:I:1}
I=\int_0^\infty \frac{dy}{y^2}\;e^{ify}\left[ u_0(y)\right] ^2,
\ee
where $f$ is a real constant. With the expression~(\ref{WF_0}) for $u_0(y)$ the integral becomes a series
\be\label{app:I:3}
I=\sum_n\sum_mc_nc_m\int_0^\infty \frac{dy}{y^2}\;e^{(if-\alpha_n-\alpha_m)y},
\ee
with all terms divergent. We regularize the expressions defining
\be\label{app:I:4}
I=\lim_{\varepsilon \to 0}I_\varepsilon,
\ee
where
\be\label{app:I:5}
\begin{split}
I_\varepsilon&=\int_0^\infty \frac{dy}{y^{2-\varepsilon}}\;e^{ify}\left[ u_0(y)\right] ^2=\\
&=\sum_n\sum_mc_nc_m(\alpha_n+\alpha_m-if)^{1-\varepsilon}\Gamma(-1+\varepsilon).
\end{split}
\ee
Expanding the $\Gamma$-function near the pole
\be\label{app:I:6}
\Gamma(-1+\varepsilon)=-\frac1{\varepsilon}+\gamma-1-\mathcal O(\varepsilon) 
\ee
and taking into account the constraint $\sum_nc_n=0$ one gets
\be\label{app:I:7}
\begin{split}
I&=\lim_{\varepsilon \to 0}\left[ \sum_n\sum_mc_nc_m(\alpha_n+\alpha_m-if)\ln(\alpha_n+\alpha_m-if)+\mathcal O(\varepsilon)\right] =\\
&=\sum_n\sum_mc_nc_m(\alpha_n+\alpha_m-if)\ln(\alpha_n+\alpha_m-if).
\end{split}
\ee
%
%


\begin{thebibliography}{99}
\bibitem{Jlab.Jones}
M.K.~Jones et al., Phys. Rev. Lett., {\bf 84}, 1398 (2000).
\bibitem{Jlab.Gayou} 
O.~Gayou et al., Phys. Rev. Lett., {\bf 88}, 092301 (2002).
\bibitem{Punjabi}
V.~Punjabi et al., Phys. Rev., C {\bf 71}, 055202 (2005); ibid., C {\bf 71}, 069902 (E) (2005).
\bibitem{Guichon}
P.A.M.~Guichon and M.~Vanderhaeghen, Phys. Rev. Lett., {\bf 91}, 142303 (2003).
\bibitem{Blunden}
P.G.~Blunden, W.~Melnitchouk, and J.A.~Tjon, Phys. Rev. Lett., {\bf 91}, 142304 (2003).
\bibitem{BorisyukKob}
D.~Borisyuk and A.~Kobushkin, Phys. Rev., C {\bf 74}, 065203 (2006).
\bibitem{BorisyukKob_phen}
D.~Borisyuk and A.~Kobushkin, Phys. Rev., C {\bf 76}, 022201 (R) (2007).
\bibitem{DeForest}
T.~De Forest, Jr., J.D.~Walecka, Adv. Phys., {\bf 15} 1 (1966). 
\bibitem{Gunion}
J.~Gunion and L.~Stodolsky, Phys. Rev. Lett., {\bf 30} 345 (1973).
\bibitem{Franco}
V.~Franco, Phys. Rev., {\bf D8} 826 (1973).
\bibitem{Boitsov}
V.N.~Boitsov, L.A.~Kondratyuk, and V.B.~Kopeliovich, Yad. Fiz., {\bf 16} 515 (1972).
\bibitem{Lev}
F.M.~Lev, Yad. Fiz., {\bf 21} 89 (1975).
\bibitem{DongChen}
Yu Bing Dong and D.Y.~Chen, Phys. Lett., {\bf B 675} 426 (2009).
\bibitem{DongPRC}
Yu Bing Dong, Chung Wen Kao, Shin Nan Yang, and Yu Chun Chen, Phys. Rev., C {\bf 74} 064006 (2006).
\bibitem{GakhTomasi}
G.I.~Gakh and E.~Tomasi-Gustaffson, Nucl. Phys., {\bf A799} 127 (2008).
\bibitem{Gourdin}
M.~Gourdin, Nuov. Cim., {\bf 28} 533 (1963).
\bibitem{KolybasovSmorodinskaya}
V.M.~Kolybasov and N.Ya.~Smorodinskaya, Phys. Lett., {\bf B 37} 272 (1971); Yad. Fiz., {\bf 17} 1211 (1973).
\bibitem{Paris}
M.~Lacombe et al., Phys. Lett., {\bf B 101} 139 (1981).
\bibitem{CD-Bonn}
R.~Machleidt, Phys. Rev., {\bf C 63} 024001 (2001).
\bibitem{NIJM}
http://nn-online.org/NN/?page=deuteronwavefunctions-table
\bibitem{PPV}
C.F.~Perdrisat, V.~Punjabi, and M.~Vanderhaeghen, Prog. Part. Nucl. Phys., {\bf 59}, 694 (2007).
\bibitem{Gastler}
S.~Gastler et al., Nucl. Phys., {\bf B32}, 221 (1971).
\end{thebibliography}
\end{document}